\documentclass[aps,preprint]{revtex4}%
\usepackage{amsmath}
\usepackage{amsfonts}
\usepackage{amssymb}
\usepackage{graphicx}
\usepackage{appendix}%
\providecommand{\U}[1]{\protect\rule{.1in}{.1in}}

\begin{document}
\title{Temperature effects on a network of dissipative quantum harmonic oscillators:
collective damping and diffusion processes}
\author{M. A. de Ponte$^{1}$, S. S. Mizrahi$^{1}$, and M. H. Y. Moussa$^{2}$.}
\affiliation{$^{1}$Departamento de F\'{\i}sica, Universidade Federal de S\~{a}o Carlos,
Caixa Postal 676, S\~{a}o Carlos, 13565-905, S\~{a}o Paulo,\textit{ }Brazil}
\affiliation{$^{2}$ Instituto de F\'{\i}sica de S\~{a}o Carlos, Universidade de S\~{a}o
Paulo, Caixa Postal 369, 13560-590 S\~{a}o Carlos, SP, Brazil }

\begin{abstract}
In this article we extend the results presented in Ref. [Phys. Rev. A
\textbf{76}, 032101 (2007)] to treat quantitatively the effects of reservoirs
at finite temperature in a bosonic dissipative network: a chain of coupled
harmonic oscillators whichever its topology, i.e., whichever the\textit{\ way}
the oscillators are coupled together, the\textit{\ strength} of their
couplings and their \textit{natural frequencies}. Starting with the case where
distinct reservoirs are considered, each one coupled to a corresponding
oscillator, we also analyze the case where a common reservoir is assigned to
the whole network. Master equations are derived for both situations and both
regimes of weak and strong coupling strengths between the network oscillators.
Solutions of these master equations are presented through the normal ordered
characteristic function. We also present a technique to estimate the
decoherence time of network states by computing separately the effects of
diffusion and the attenuation of the interference terms of the Wigner
function. A detailed analysis of the diffusion mechanism is also presented
through the evolution of the Wigner function. The interesting collective
diffusion effects are discussed and applied to the analysis of decoherence of
a class of network states. Finally, the entropy and the entanglement of a pure
bipartite system are discussed.

\end{abstract}

\maketitle

\section{Introduction}

The subject of networks of interacting quantum systems has acquired an
important role in the emerging field of quantum information theory. Since a
realistic quantum logic processor must ultimately be composed of a large
number of interacting quantum systems, it becomes mandatory to understand
processes as like as perfect state transfer from one to another system of the
network, and even to compute the fidelity of such a state transfer when the
action of the environment is taken into account. A significant amount of
result has recently been derived on the subject of perfect state transfer in
optical lattices \cite{Feder} and networks of spin \cite{Spin} and harmonic
oscillators \cite{Plenio}. Perfect state transfer has also been considered in
networks of arbitrary topology and coupling configuration \cite{Topology} and
even under random fluctuations in the couplings of a quantum chains
\cite{Bose}.

Apart from state transfer, the process of decoherence of a network state has
also attracted attention and interesting properties of collective damping
effects, as the nonadditivity of decoherence rates, have been discussed in
different contexts as in superconducting qubits \cite{Brito}, two-atom systems
\cite{Ficek}, and chains of dissipative harmonic oscillators
\cite{Mickel1,Mickel2,MickelDFS,MickelGeral}. Still regarding the process of
collective decoherence, the emergence of decoherence-free subspaces (DFSs) has
also instigated several interesting results when considering the particular
case of a composite system interacting with a common reservoir \cite{ZR}, or
the more realistic situation where each system interacts with its own
reservoir \cite{MickelDFS}. We call the attention to the fact that all Refs.
\cite{Mickel1,Mickel2,MickelDFS} envisage such realistic cases of networks
where each oscillator interacts with its own reservoir, also addressing the
particular case where a common reservoir is considered.

To better understand the results in Refs. \cite{Mickel1,Mickel2,MickelDFS},
which are crucial to introduce the subject of the present work, we remember
that, apart from the distinct reservoirs, the network of $N$ dissipative
harmonic oscillators could present direct an indirect dissipative channels.
Through the direct channels each oscillator loses excitation to its own
reservoir, whereas through the indirect channels it loses excitation to all
the other reservoirs but to its own. When we consider distinct reservoirs for
each network oscillator, the indirect dissipative channels --- intrinsically
associated with the nonadditivity of decoherence rates and the emergence of
DFSs \cite{MickelDFS} --- are significant only in the strong coupling regime
where $N\lambda_{mn}\simeq\omega_{\ell}$, i.e., the number of network
oscillators $N$ multiplied by their coupling strengths $\left\{  \lambda
_{mn}\right\}  $ are about their natural frequencies $\left\{  \omega
_{m}\right\}  $. Therefore, the strong coupling regime, which brings together
the collective damping effects, depends on the number of network oscillators
as much as on their coupling strengths. For Markovian white noise reservoirs,
however, where the spectral densities of the reservoirs are invariant over
translation in frequency space, the indirect channels becomes null, except for
the case $N=2$ \cite{Mickel1}.

In the weak coupling regime where $N\lambda_{mn}\ll\omega_{\ell}$, the
indirect channels always disappears. However, these indirect channels, coming
from the strong coupling regime, remains in the case where all network
oscillators interacts with a common reservoir \cite{MickelDFS}, even assuming
a common Markovian white noise reservoir. This is due to the fact that a
common reservoir induces an additional correlation between the network
oscillators, restoring the indirect decay channels.

Recently, a generalization of Refs. \cite{Mickel1,Mickel2} has been presented
through a comprehensive treatment of networks of dissipative quantum harmonic
oscillators, whichever its topology, i.e., whichever the\textit{\ way} the
oscillators are coupled together, the\textit{\ strength} of their couplings
and their \textit{natural frequencies }\cite{MickelGeral}. Focusing on the
general more realistic scenario where each oscillator is coupled to its own
reservoir, the case where all the network oscillators are coupled to a common
reservoir was also addressed. However, after deducing the master equation for
the case where all the reservoirs are at finite temperatures, all further
analysis of the dynamics of the network states was restricted to the case
where the reservoirs are at $T=0$ K. Whereas a quantitative analysis of the
decoherence and the evolution of the linear entropy of representative states
of the network were given at $0$ K, only a brief qualitative analysis of the
equilibrium states of the network was presented at finite temperatures. In the
present manuscript we extend the treatment in Ref. \cite{MickelGeral} given a
detailed analysis of the temperature effects on networks of dissipative
quantum harmonic oscillators.

The present extension of Ref. \cite{MickelGeral} that accounts for the
temperature effects coming from thermal reservoirs is not only interesting due
to its more realistic approach but also from the mathematical development here
achieved. In fact, we present an alternative approach to previous results in
the literature \cite{Gardiner} regarding the obtainment of the solution of the
master equation and the estimation of decoherence times through the Wigner
distribution function. To circumvent noise effects, many of the nowadays
experiments demonstrating quantum logic operations through atom-field
interactions occur in cryogenic environments where temperature effects are
negligible. In cavity quantum electrodynamics, the setup is cooled to around
0.5 K by a $^{3}$He-$^{4}$He refrigerator to avoid blackbody radiation in the
High-Q superconducting cavity. Under such a specific condition, the
temperature effects on the decoherence process are almost negligible. However,
when the setup is scaled from one single cavity to a network of $N$ High-Q
cavities, major questions arise due to temperature effects. First of all,
would the DFSs survive despite the temperature effects? Apart from the special
class of states composing the DFSs, how the temperature affects other states
of the network, as for example initial entangled states? Evidently, these
questions present no obvious answers, even under the assumption that all
network cavities are cooled at low temperatures. On this regard, we expect
that the collective damping effects coming from the indirect dissipative
channels to play a major role for the answers to the above questions.

Apart from providing the mathematical treatment of the temperature effects on
a network of $N$ dissipative harmonic oscillators, in the present manuscript
we also analyze the role played by temperature on the evolution of particular
states of the network other than those composing DFSs. We reserve the analyses
of the emergence of DFSs under temperature effects to an specific work
\cite{Mickel4} where the mechanism of construction of such privileged states
is also discussed along with decoherence.

This paper is organized as following: in Section II we revisit our model of a
bosonic dissipative network \cite{MickelGeral} and present the derivation of
the master equation governing the dynamics of the associated density operator.
In Section III we present the solution of the normal ordered characteristic
equation obtained from the master equation for the density operator of the
network. In Section IV we analyze the evolution of two general classes of
initial states of the network, given by mixtures of coherent and number
states, through the normal ordered characteristic equation, the
Glauber-Sudarshan $P$-function, and the Wigner distribution. A detailed
analyses of the diffusion processes is presented in Section V and the
collective decoherence rates of a family of states of the network is analyzed
in Section VI. In Section VII we discuss the entropy and the entanglement
degree of a pure bipartite system and, finally, in Section VIII we present our
concluding remarks.

\section{The master equation of a bosonic dissipative network}

We present here a brief review of the steps for the derivation of the master
equation of a bosonic network, as developed in Ref. \cite{MickelGeral}. We
start from the general case of a network of $N$ interacting oscillators, where
each one interacts with each other, from which all other topologies can be
recovered. As depicted in Fig. 1, we also consider the case where each
oscillator interacts with its own reservoir due to this more realistic
approach for most of the physical systems. However, as pointed in Ref.
\cite{MickelDFS}, despite the realistic scenario of the case of distinct
reservoirs, the case of a common reservoir is more general from the technical
point of view. In fact, as discussed at the end of this section, the master
equation for the case of distinct reservoirs can be deduced from the case of a
common reservoir.

We start by considering a general Hamiltonian for a bosonic network,
$H=H_{S}+H_{R}+H_{I}$, composed by a set of $N$ coupled oscillators
\begin{equation}
H_{S}=\hbar\sum_{m=1}^{N}\left[  \omega_{m}a_{m}^{\dag}a_{m}+\frac{1}{2}%
\sum_{n\left(  \neq m\right)  =1}^{N}\lambda_{mn}\left(  a_{m}^{\dag}%
a_{n}+a_{m}a_{n}^{\dag}\right)  \right]  \text{,} \label{1}%
\end{equation}
$N$ distinct reservoirs, modeled by a set of $k=1,\ldots,\infty$ modes,%
\begin{equation}
H_{R}=\hbar\sum_{m=1}^{N}\sum_{k}\omega_{mk}b_{mk}^{\dag}b_{mk}\text{,}
\label{2}%
\end{equation}
and the coupling between the network oscillators and their respective
reservoirs%
\begin{equation}
H_{I}=\hbar\sum_{m=1}^{N}\sum_{k}V_{mk}\left(  b_{mk}^{\dag}a_{m}+b_{mk}%
a_{m}^{\dag}\right)  \text{,} \label{3}%
\end{equation}
where $b_{mk}^{\dagger}$ ($b_{mk}$) is the creation (annihilation) operator
for the $k$th bath mode $\omega_{mk}$ coupled to the $m$th network oscillator
$\omega_{m}$ whose creation (annihilation) operator reads $a_{m}^{\dagger}$
($a_{m}$). The coupling strengths between the oscillators are given by the set
$\left\{  \lambda_{mn}\right\}  $, while those between the oscillators and
their reservoirs by $\left\{  V_{mk}\right\}  $. We assume, from here on, that
$\ell,m,n,\ell^{\prime},m^{\prime},$ and $n^{\prime}$ run from $1$ to $N$.

Before addressing the dissipative process through Hamiltonian (\ref{3}), we
focus first on Hamiltonian $H_{S}$ to show how to derive different topologies
of a nondissipative network of coupled harmonic oscillators. Rewriting $H_{S}$
in a matrix form%
\begin{equation}
H_{S}=\hbar\left(
\begin{array}
[c]{ccc}%
a_{1}^{\dag} & \cdots & a_{N}^{\dag}%
\end{array}
\right)  \left(
\begin{array}
[c]{ccc}%
\mathcal{H}_{11} & \cdots & \mathcal{H}_{1N}\\
\vdots & \ddots & \vdots\\
\mathcal{H}_{N1} & \cdots & \mathcal{H}_{NN}%
\end{array}
\right)  \left(
\begin{array}
[c]{c}%
a_{1}\\
\vdots\\
a_{N}%
\end{array}
\right)  \text{,} \label{3.5}%
\end{equation}
we identify the elements of the matrix $\mathcal{H}=\mathcal{H}^{\dag}$ as%
\begin{equation}
\mathcal{H}_{mn}=\left\{
\begin{array}
[c]{ccc}%
\omega_{m} & \text{for} & m=n\\
\lambda_{mn} & \text{for} & m\neq n
\end{array}
\right.  \text{,} \label{4}%
\end{equation}
whose values characterize whichever the network topology, i.e., whichever the
way the oscillators are coupled together, the set of coupling strengths
$\left\{  \lambda_{mn}\right\}  $, and their natural frequencies $\left\{
\omega_{m}\right\}  $.

To obtain the master equation of the network we first diagonalize the
Hamiltonian $\mathcal{H}$ (within the physical regime where the normal modes
are assumed to be positive) through the canonical transformation%
\begin{equation}
A_{m}=\sum_{n}C_{mn}a_{n}\text{,} \label{5}%
\end{equation}
where the coefficients of the $m$th line of matrix $\mathbf{C}$ define the
eigenvectors associated to the eigenvalues $\varpi_{m}$ of matrix
$\mathcal{H}$. With $\mathbf{C}$ being an orthogonal matrix, its transposed
$\mathbf{C}^{\intercal}$ turns out to be exactly\ its inverse $\mathbf{C}%
^{-1}$, resulting in the commutation relations $\left[  A_{m},A_{n}^{\dag
}\right]  =\delta_{mn}$ and $\left[  A_{m},A_{n}\right]  =0$, which enable the
Hamiltonian $H$ to be rewritten as a sum $H=H_{0}+V$, where
\begin{subequations}
\label{6}%
\begin{align}
H_{0}  &  =\hbar\sum_{m}\varpi_{m}A_{m}^{\dag}A_{m}+\hbar\sum_{m}\sum
_{k}\omega_{mk}b_{mk}^{\dag}b_{mk}\text{,}\label{6a}\\
V  &  =\hbar\sum_{m,n}\sum_{k}C_{mn}^{-1}V_{mk}\left(  b_{mk}^{\dag}%
A_{n}+b_{mk}A_{n}^{\dag}\right)  \text{.} \label{6b}%
\end{align}
With the diagonalized Hamiltonian $H_{0}$ we are ready to introduce the
interaction picture, defined by the transformation $U_{0}(t)=\exp\left(
-iH_{0}t/\hbar\right)  $, where
\end{subequations}
\begin{equation}
V(t)=\hbar\sum_{m,n}\left[  \mathcal{O}_{mn}(t)A_{n}^{\dag}+\mathcal{O}%
_{mn}^{\dag}(t)A_{n}\right]  \text{,} \label{7}%
\end{equation}
and $\mathcal{O}_{mn}(t)=C_{mn}^{-1}\sum_{k}V_{mk}\exp\left[  -i\left(
\omega_{mk}-\varpi_{n}\right)  t\right]  b_{mk}$. Next, we assume that the
interactions between the resonators and the reservoirs are weak enough to
allow a second-order perturbation approximation. We also assume a Markovian
reservoir such that the density operator of the global system can be
factorized as $\rho_{S}(t)\otimes\rho_{R}(0)$. Under these assumptions the
reduced density operator of the network of $N$ dissipative coupled resonators
satisfy the differential equation%
\begin{equation}
\frac{d\rho_{S}(t)}{dt}=-\frac{1}{\hbar^{2}}\int_{0}^{t}d\tau
\operatorname*{Tr}\nolimits_{R}\left[  V(t),\left[  V(\tau),\rho_{S}%
(t)\otimes\rho_{R}(0)\right]  \right]  \text{.} \label{8}%
\end{equation}
Since for a thermal reservoir $\left\langle b_{mk}b_{nk^{\prime}}\right\rangle
=\left\langle b_{mk}^{\dag}b_{nk^{\prime}}^{\dag}\right\rangle =0$, we have to
solve the integrals appearing in Eq. (\ref{8}), related to correlation
functions of the form%
\begin{align}
\int_{0}^{t}d\tau\left\langle \mathcal{O}_{mn}(t)\mathcal{O}_{m^{\prime}\ell
}^{\dag}(\tau)\right\rangle  &  =C_{mn}^{-1}C_{\ell m^{\prime}}\int_{0}%
^{t}d\tau\sum_{k,k^{\prime}}V_{mk}V_{m^{\prime}k^{\prime}}\left\langle
b_{mk}b_{m^{\prime}k^{\prime}}^{\dag}\right\rangle \nonumber\\
&  \times\exp\left\{  -i\left[  \left(  \omega_{mk}-\varpi_{n}\right)
t-\left(  \omega_{m^{\prime}k^{\prime}}-\varpi_{\ell}\right)  \tau\right]
\right\}  . \label{9}%
\end{align}
Considering that the reservoir frequencies are very closely spaced to allow a
continuum summation, we obtain
\begin{equation}
\int_{0}^{t}d\tau\left\langle \mathcal{O}_{mn}(t)\mathcal{O}_{m^{\prime}\ell
}^{\dag}(\tau)\right\rangle =N\delta_{mm^{\prime}}C_{mn}^{-1}C_{\ell m}%
\frac{\gamma_{m}(\varpi_{\ell})+\bar{n}_{m}(\varpi_{\ell})\tilde{\gamma}%
_{m}(\varpi_{\ell})}{2}\operatorname*{e}\nolimits^{i\left(  \varpi_{\ell
}-\varpi_{n}\right)  t}\text{,} \label{10}%
\end{equation}
where we have defined the average excitation of the reservoir associated to
the $m$th oscillator as $\bar{n}_{m}\left(  \nu\right)  $ through the relation
$\left\langle b_{m}^{\dagger}(\nu)b_{n}(\nu^{\prime})\right\rangle =2\pi
\delta_{mn}\bar{n}_{m}(\nu)\delta\left(  \nu-\nu^{\prime}\right)  $, apart
from the damping rates
\begin{subequations}
\label{11}%
\begin{align}
\gamma_{m}(\omega)  &  =\int_{0}^{t}d\tau\int_{0}^{\infty}\frac{d\nu}{N\pi
}\left[  V_{m}(\nu)\sigma_{m}(\nu)\right]  ^{2}e^{-i\left(  \nu-\omega\right)
\left(  \tau-t\right)  }\text{,}\label{11a}\\
\tilde{\gamma}_{m}(\omega)  &  =\int_{0}^{t}d\tau\int_{0}^{\infty}\frac{d\nu
}{N\pi}\left[  V_{m}(\nu)\sigma_{m}(\nu)\right]  ^{2}\frac{\bar{n}_{m}(\nu
)}{\bar{n}_{m}(\omega)}e^{-i\left(  \nu-\omega\right)  \left(  \tau-t\right)
}\text{,} \label{11b}%
\end{align}
with $\sigma_{m}(\nu)$ being the density of states of the $m$th reservoir. In
the context of Markov approximation, where $V_{m}(\varpi_{n}),\sigma
_{m}(\varpi_{n})$ and $\bar{n}_{m}(\varpi_{n})$ are slowly varying functions
around the normal modes $\varpi_{n}$ we can simplify the expressions
(\ref{11}) to their usual forms%
\end{subequations}
\begin{equation}
\gamma_{m}(\omega)=\tilde{\gamma}_{m}(\omega)=\frac{1}{N}\left[  V_{m}%
(\omega)\sigma_{m}(\omega)\right]  ^{2}\text{.} \label{11.5}%
\end{equation}

Back to the Schr\"{o}dinger picture and to the original field operators
$a_{m}$, we finally obtain from the steps outlined above, the master equation%
\begin{equation}
\frac{d\rho_{S}(t)}{dt}=\frac{i}{\hbar}\left[  \rho_{S}(t),H_{S}\right]
+\sum_{m,n}\left[  \frac{\Gamma_{mn}+\Upsilon_{mn}}{2}\mathcal{L}_{mn}\rho
_{S}(t)+\frac{\Upsilon_{mn}}{2}\mathfrak{L}_{mn}\rho_{S}(t)\right]  \text{,}
\label{12}%
\end{equation}
where we have defined the damping and the diffusion matrix elements
$\Gamma_{mn}$ and $\Upsilon_{mn}$ in the forms
\begin{subequations}
\label{13}%
\begin{align}
\Gamma_{mn}  &  =N\sum_{\ell}C_{\ell n}\gamma_{m}(\varpi_{\ell})C_{m\ell}%
^{-1}\text{,}\label{13a}\\
\Upsilon_{mn}  &  =N\sum_{\ell}C_{\ell n}\tilde{\gamma}_{m}(\varpi_{\ell}%
)\bar{n}_{m}(\varpi_{\ell})C_{m\ell}^{-1}\text{,} \label{13b}%
\end{align}
whereas the Liouville operators accounting for the \emph{direct} ($m=n$) and
\emph{indirect} ($m\neq n$)\ dissipative channels, are given by
\end{subequations}
\begin{subequations}
\label{14}%
\begin{align}
\mathcal{L}_{mn}\rho_{S}(t)  &  \equiv\left[  a_{n}\rho_{S}(t),a_{m}^{\dag
}\right]  +\left[  a_{m},\rho_{S}(t)a_{n}^{\dag}\right]  ,\label{14a}\\
\mathfrak{L}_{mn}\rho_{S}(t)  &  \equiv\left[  a_{n}^{\dag}\rho_{S}%
(t),a_{m}\right]  +\left[  a_{m}^{\dag},\rho_{S}(t)a_{n}\right]  . \label{14b}%
\end{align}

As mentioned in the Introduction and discussed in Ref. \cite{MickelGeral}, the
oscillators lose excitation to their own reservoirs through the direct
dissipative channels, whereas through the indirect channels they lose
excitation to all the other reservoirs but not to their own. Although in Ref.
\cite{MickelGeral} we have obtained the master equation for the general case
of reservoirs at finite temperatures, all further analysis was carried out for
reservoirs at $0$ K where the diffusion matrix elements $\Upsilon_{mn}$ are
null. Next, considering the case of reservoirs at finite temperatures, we must
discuss the master equation (\ref{12}) under the weak and strong coupling
regimes between the network oscillators.

\subsection{Weak coupling regime}

We first remember that the weak coupling regime for a network of $N$ coupled
oscillator is defined by the relation $N\lambda_{mn}\ll\omega_{\ell}$.
(However, if a specific coupling $\lambda_{mn}$ between two oscillators, $m$
and $n$, fails to satisfy the relation $N\lambda_{mn}\ll\omega_{\ell}$, the
network dynamics is necessarily described by the strong coupling regime, with
some normal-mode frequencies far beyond their natural values.) In the weak
coupling regime, the interaction between the network oscillators, described by
$\hbar\sum_{m,\neq n}\lambda_{mn}\left(  a_{m}^{\dag}a_{n}+a_{m}a_{n}^{\dag
}\right)  /2$, could be directly introduced into the von Neumann term of the
master equation to a good approximation, circumventing the necessity to
diagonalize the Hamiltonian $\mathcal{H}$ through a canonical transformation
$A_{m}=\sum_{n}C_{mn}a_{n}$. This is equivalent to approximate the matrix
$\mathbf{C}$ by a identity matrix $\mathbf{I}$, implying that
\end{subequations}
\begin{subequations}
\label{15}%
\begin{align}
\Gamma_{mn}  &  =N\gamma_{m}(\omega_{m})\delta_{mn}\text{,}\label{15a}\\
\Upsilon_{mn}  &  =N\tilde{\gamma}_{m}(\omega_{m})\bar{n}_{m}(\omega
_{m})\delta_{mn}\text{,} \label{15b}%
\end{align}
where, evidently, we have also approximated the normal modes by their original
natural frequencies. Under the above considerations, the master equation
(\ref{12}) becomes
\end{subequations}
\begin{align}
\frac{d\rho_{S}(t)}{dt}  &  =\frac{i}{\hbar}\left[  \rho_{S}(t),H_{S}\right]
+N\sum_{m}\left[  \frac{\gamma_{m}(\omega_{m})+\tilde{\gamma}_{m}(\omega
_{m})\bar{n}_{m}(\omega_{m})}{2}\mathcal{L}_{mm}\rho_{S}(t)\right. \nonumber\\
&  +\left.  \frac{\tilde{\gamma}_{m}(\omega_{m})\bar{n}_{m}(\omega_{m})}%
{2}\mathfrak{L}_{mm}\rho_{S}(t)\right]  \text{,} \label{16}%
\end{align}
where, essentially, the indirect dissipative channels disappear, establishing
the additivity of the decoherence rates.

\subsection{Strong coupling regime}

The strong coupling regime means that $N\lambda_{mn}\approx\omega_{\ell}$,
i.e., at least one of the coupling $\left\{  \lambda_{mn}\right\}  $ between
two network oscillators must be of the order of any natural frequency
$\left\{  \omega_{m}\right\}  $. In this case, the indirect dissipative
channels become effective, inducing collective damping and diffusion effects
that we must investigate. As pondered in the Introduction, would the
collective effects of nonadditivity of the decay rates and the emergence of
DFSs still survive despite the temperature effects?

It must be mentioned that Markovian white noise reservoirs washes out the
collective damping effects introduced by the strong coupling regime since the
spectral densities are invariant over translation in frequency space, i.e.,
$\gamma_{m}(\varpi_{n})=\gamma_{m}$, rendering the same matrix elements
$\Gamma_{mn}$ as in Eq. (\ref{15a}). However, Markovian white noise reservoirs
do not washes the collective diffusion effects. In fact, only under the
\textit{additional} assumption that $\bar{n}_{m}(\varpi_{n})\approx\bar{n}%
_{m}$ for whatever normal mode $\varpi_{n}$, we recover Eq. (\ref{15b}) for
the diffusion matrix $\mathbf{\Upsilon}$, erasing the collective effects
completely. Next we discuss the case where the whole network is under the
action of a common reservoir.

\subsection{A common reservoir}

When all the network oscillators are coupled to a single reservoir, the master
equation, derived in Ref. \cite{MickelDFS}, is similar to that in Eq.
(\ref{12}), replacing the damping and the diffusion matrix elements by%
\begin{subequations}
\begin{align}
\Gamma_{mn}  &  =N\sum_{\ell,n^{\prime}}C_{\ell n}\gamma_{mn^{\prime}}\left(
\varpi_{\ell}\right)  C_{n^{\prime}\ell}^{-1},\label{17a}\\
\Upsilon_{mn}  &  =N\sum_{\ell,n^{\prime}}C_{\ell n}\tilde{\gamma}%
_{mn^{\prime}}\left(  \varpi_{\ell}\right)  C_{n^{\prime}\ell}^{-1}\text{,}
\label{17b}%
\end{align}
where the damping rates $\gamma_{mn}(\omega)$ and $\tilde{\gamma}_{mn}\left(
\omega\right)  $ for the case of a single common reservoir are given by
\cite{MickelDFS}
\end{subequations}
\begin{subequations}
\label{18}%
\begin{align}
\gamma_{mn}\left(  \omega\right)   &  =\int_{0}^{t}d\tau\int_{0}^{\infty}%
\frac{d\nu}{N\pi}V_{m}\left(  \nu\right)  V_{n}\left(  \nu\right)  \sigma
^{2}(\nu)e^{-i\left(  \nu-\omega\right)  \tau},\label{18a}\\
\tilde{\gamma}_{mn}\left(  \omega\right)   &  =\int_{0}^{t}d\tau\int
_{0}^{\infty}\frac{d\nu}{N\pi}V_{m}\left(  \nu\right)  V_{n}\left(
\nu\right)  \sigma^{2}(\nu)\frac{\bar{n}\left(  \nu\right)  }{\bar{n}(\omega
)}e^{-i\left(  \nu-\omega\right)  \tau}. \label{18b}%
\end{align}

As mentioned above and discussed in Ref. \cite{MickelDFS}, the master equation
for the case of distinct reservoirs can be deduced from the case of a single
common reservoir. The above deduction of the master equation (\ref{12}), where
we started from the case of distinct reservoirs, was entirely due to its broad
application in many physical systems. To demonstrate how to derive the case of
distinct reservoir from that of a common one, we remember that $V_{m}\left(
\nu\right)  $ gives the distribution function of the reservoir modes coupled
to the $m$th oscillator. Therefore, in the absence of overlap between the
distribution functions, i.e., $\int d\nu V_{m}(\nu)V_{n}(\nu)=0$ for $m\neq
n$, Eqs. (\ref{18}) reduce to those in Eqs. (\ref{11}). In this case, the
occurrence of the indirect-decay channels follows entirely from the strong
coupling between the oscillators, as discussed in the subsections presented
above. When there is a significant overlap between the distribution functions,
i.e., $\int d\nu V_{m}(\nu)V_{n}(\nu)\neq$ $0$ for at least one $m\neq n$, we
get the indirect-decay channels even when the network oscillators do not
interacts at all. The strength of the damping and the diffusion matrix
elements being defined by the amount of the overlap, i.e., when the overlap
between the distributions $V_{m}(\nu)$ and $V_{n}(\nu)$ is maximum, the
strengths $\Gamma_{mn}$ and $\Upsilon_{mn}$ equals $\Gamma_{mm}$ and
$\Upsilon_{mm}$.

\section{Normal ordered characteristic function}

To analyze the dynamics of the network states for the case where the
reservoirs are at finite temperatures, we consider the evolution of the
(normal ordered) characteristic function, derived from the master equation
(\ref{12}) (suitable for all cases discussed in the previous Section) as%
\end{subequations}
\begin{equation}
\frac{d}{dt}\chi(\{\eta_{m}\},t)=-\sum_{m,n}\left[  \eta_{m}\frac
{\Upsilon_{mn}}{2}\eta_{n}^{\ast}+\eta_{m}\left(  \mathcal{H}_{mn}^{D}\right)
^{\ast}\frac{\partial}{\partial\eta_{n}}+c.c.\right]  \chi(\{\eta
_{m}\},t)\text{,} \label{19}%
\end{equation}
where we defined the matrix elements
\begin{equation}
\mathcal{H}_{mn}^{D}=\Gamma_{mn}/2+i\mathcal{H}_{mn}\text{.} \label{20}%
\end{equation}
As noted in Ref. \cite{MickelGeral}, the matrix $\mathcal{H}^{D}$ is an
extension of the free evolution $\mathcal{H}$ in Eq. (\ref{4}), which takes
into account the dissipative mechanisms of the network.

Starting with the assumption that Eq. (\ref{19}) admits a solution of the form
$\chi(\{\eta_{m}\},t)=\varphi(\{\eta_{m}\})\phi(\{\eta_{m}\},t)$, we obtain
two differential equations, one accounting for the dynamic process, given by%
\begin{equation}
\frac{d}{dt}\phi(\{\eta_{m}\},t)=-\sum_{m,n}\left[  \eta_{m}\left(
\mathcal{H}_{mn}^{D}\right)  ^{\ast}\frac{\partial}{\partial\eta_{n}%
}+c.c.\right]  \phi(\{\eta_{m}\},t)\text{,} \label{21}%
\end{equation}
and the other accounting for the stationary solution of the characteristic
function, given by%
\begin{equation}
\sum_{m,n}\left[  \eta_{m}\frac{\Upsilon_{mn}}{2}\eta_{n}^{\ast}+\eta
_{m}\left(  \mathcal{H}_{mn}^{D}\right)  ^{\ast}\frac{\partial}{\partial
\eta_{n}}+c.c.\right]  \varphi(\{\eta_{m}\})=0\text{.} \label{22}%
\end{equation}
If we perform the substitution $\mathcal{H}^{D}\rightarrow-\left(
\mathcal{H}^{D}\right)  ^{\dagger}$ in the first differential equation
(\ref{21}), it turns out to be exactly that appearing in Ref.
\cite{MickelGeral} for the derivation of the solution of the Glauber-Sudarshan
$P$-function. Therefore, following the steps outlined in Ref.
\cite{MickelGeral}, the solution of Eq. (\ref{21}) can be written as%
\begin{equation}
\eta_{m}(t)=\sum_{\ell,n}\eta_{n}\left(  0\right)  D_{n\ell}^{\ast}\exp\left(
-\Omega_{\ell}^{\ast}t\right)  \left(  D_{\ell m}^{-1}\right)  ^{\ast}\text{,}
\label{23}%
\end{equation}
where we employed the diagonal form of $\mathcal{H}^{D}$ following from the
transformation $\mathbf{D}^{-1}\bullet\mathcal{H}^{D}\bullet\mathbf{D=\Omega}%
$. Note that writing the solution (\ref{23}) in a matrix form, it becomes%
\begin{align}
\mathbf{\eta}(t)  &  =\mathbf{\eta}\left(  0\right)  \bullet\mathbf{D}^{\ast
}\bullet\exp\left(  -\mathbf{\Omega}^{\ast}t\right)  \bullet\left(
\mathbf{D}^{-1}\right)  ^{\ast}\nonumber\\
&  \mathbf{=\eta}\left(  0\right)  \bullet\exp\left[  -\left(
\mathbf{D\bullet\Omega\bullet D}^{-1}\right)  ^{\ast}t\right] \nonumber\\
&  =\mathbf{\eta}\left(  0\right)  \bullet\exp\left[  -\left(  \mathcal{H}%
^{D}\right)  ^{\ast}t\right]  \label{24}%
\end{align}
such that%
\begin{equation}
\frac{d\mathbf{\eta}\left(  t\right)  }{dt}=-\mathbf{\eta}\left(  t\right)
\bullet\left(  \mathcal{H}^{D}\right)  ^{\ast}\text{,} \label{25}%
\end{equation}
or, equivalently,
\begin{equation}
\frac{d\eta_{n}(t)}{dt}=-\sum_{m}\eta_{m}(t)\left(  \mathcal{H}_{mn}%
^{D}\right)  ^{\ast}\text{,} \label{26}%
\end{equation}
representing a system of coupled differential equations which follows from Eq.
(\ref{21}) under the assumption that $\phi(\{\eta_{m}\},t)=\phi(\{\eta
_{m}(t)\})$, with $\eta_{m}=\eta_{m}(0)$ \cite{MickelGeral}.

The second differential equation (\ref{22}) can be solved assuming a general
Gaussian form%
\begin{equation}
\varphi(\{\eta_{m}\})=\exp\left(  -\frac{1}{2}\sum_{m,n}\eta_{m}\Pi_{mn}%
\eta_{n}^{\ast}\right)  \text{,}\label{27}%
\end{equation}
where the elements of matrix $\mathbf{\Pi}$ are the coefficients to be
determined. Substituting (\ref{27}) into Eq.(\ref{22}) and changing
conveniently the labels $m$ and $n$ of the involved matrices, we verify that
the differential equation (\ref{22}) reduces to a matrix equation of the form
\begin{equation}
\left(  \mathcal{H}^{D}\right)  ^{\ast}\bullet\mathbf{\Pi+\Pi}\bullet\left(
\mathcal{H}^{D}\right)  ^{\top}=\mathbf{\Upsilon}+\mathbf{\Upsilon}^{\top
}\text{,}\label{28}%
\end{equation}
which is explicitly written as%
\begin{equation}
\sum_{\ell}\left(  \mathcal{H}_{m\ell}^{D}\right)  ^{\ast}\Pi_{\ell n}%
+\sum_{\ell}\Pi_{m\ell}\mathcal{H}_{n\ell}^{D}=\Upsilon_{mn}+\Upsilon
_{nm}\text{.}\label{29}%
\end{equation}
with the superscript $\top$ in Eq. (\ref{28}) standing for transposed. It is
worth noting that for identical reservoirs, where $\gamma_{m}=\gamma$ and so
$\mathbf{\Gamma}=\mathbf{\Gamma}^{\top}$, we obtain a symmetric dissipative
matrix $\mathcal{H}^{D}$, i.e., $\mathcal{H}^{D}=\left(  \mathcal{H}%
^{D}\right)  ^{\top}$, making Eq. (\ref{28}) the well-known Lyapunov equation.
The solution of Eq. (\ref{28}), namely the determination of $\mathbf{\Pi}$,
can be obtained by converting the matrix equation into a system of $N^{2}$
algebraic equations, i.e., into a new matrix equation of the simplified form
$\mathbf{A\bullet X=B}$, with the elements of matrix $\mathbf{X}$ being the
$N^{2}$ unknown variables. To this end, it is useful to define the column
vector%
\begin{equation}
\operatorname{vec}\left(  \Pi\right)  \equiv\left(
\begin{array}
[c]{ccccccccccc}%
\Pi_{11} & \Pi_{21} & \cdots & \Pi_{N1} & \Pi_{12} & \cdots & \Pi_{N2} &
\cdots & \Pi_{1N} & \cdots & \Pi_{NN}%
\end{array}
\right)  ^{\top},\label{30}%
\end{equation}
where the first $N$ elements of $\operatorname{vec}\left(  \Pi\right)  $
correspond to the first column of matrix $\mathbf{\Pi}$, whereas the next $N$
elements correspond to the second column of $\mathbf{\Pi}$ and so on. As so,
the matrix equation (\ref{28}) can be rewritten into the form
\cite{livroSalomon}
\begin{equation}
\left[  \mathbf{I}\otimes\left(  \mathcal{H}^{D}\right)  ^{\ast}%
+\mathcal{H}^{D}\otimes\mathbf{I}\right]  \bullet\operatorname{vec}\left(
\Pi\right)  =\operatorname{vec}\left(  \mathbf{\Upsilon}+\mathbf{\Upsilon
}^{\top}\right)  \text{,}\label{31}%
\end{equation}
where $\mathbf{I}$ is an $N\times N$ identity matrix. From the mathematical
properties presented in Appendix A for the matrix $\left[  \mathbf{I}%
\otimes\left(  \mathcal{H}^{D}\right)  ^{\ast}+\mathcal{H}^{D}\otimes
\mathbf{I}\right]  $, we verify that the elements of matrix $\mathbf{\Pi}$ can
be written as%
\begin{equation}
\Pi_{\ell\ell^{\prime}}=\sum_{m,n,m^{\prime},n^{\prime}}\frac{\Upsilon
_{m^{\prime}n^{\prime}}+\Upsilon_{n^{\prime}m^{\prime}}}{\Omega_{m}+\Omega
_{n}^{\ast}}D_{\ell^{\prime}m}D_{mm^{\prime}}^{-1}\left(  D_{\ell
n}D_{nn^{\prime}}^{-1}\right)  ^{\ast}\text{,}\label{32}%
\end{equation}
finally leading to the solution of Eq. (\ref{22}) through Eq.(\ref{27}). In
fact, substituting the expression (\ref{32}) into the left hand side of Eq.
(\ref{29}), and using the relation $\mathbf{D}^{-1}\bullet\mathcal{H}%
^{D}\bullet\mathbf{D=\Omega}\Rightarrow\mathcal{H}^{D}\bullet
\mathbf{D=D\bullet\Omega}$, we obtain%
\begin{equation}
\sum_{\ell}\left[  \left(  \mathcal{H}_{m\ell}^{D}\right)  ^{\ast}\Pi_{\ell
n}+\Pi_{m\ell}\mathcal{H}_{n\ell}^{D}\right]  =\sum_{m^{\prime},n^{\prime}%
}\left(  \Upsilon_{m^{\prime}n^{\prime}}+\Upsilon_{n^{\prime}m^{\prime}%
}\right)  \delta_{mn^{\prime}}\delta_{nm^{\prime}}=\Upsilon_{mn}+\Upsilon
_{nm},\label{33}%
\end{equation}
which is exactly the right hand side of Eq. (\ref{29}).

We thus verify that the solution of the characteristic equation is of the form%
\begin{equation}
\chi(\{\eta_{m}\},t)=\varphi(\{\eta_{m}\})\left[  \phi(\{\eta_{m}%
\},t=0)\right\vert _{\left\{  \eta_{m}\right\}  \Rightarrow\left\{  \eta
_{m}(t)\right\}  }\text{.} \label{34}%
\end{equation}
Since for $t=0$ we get $\chi(\{\eta_{m}\},0)=\varphi(\{\eta_{m}\})\phi
(\{\eta_{m}\},0)$, such that $\phi(\{\eta_{m}\},0)=\chi(\{\eta_{m}%
\},0)/\varphi(\{\eta_{m}\})$, we end up with the solution of the
characteristic function
\begin{equation}
\chi(\{\eta_{m}\},t)=\frac{\varphi(\{\eta_{m}\})}{\varphi(\{\eta_{m}%
(t)\})}\left[  \chi(\{\eta_{m}\},t=0)\right\vert _{\left\{  \eta_{m}\right\}
\Rightarrow\left\{  \eta_{m}(t)\right\}  }\text{,} \label{35}%
\end{equation}
given in terms of its initial state. An interesting point to be noted is that
the dynamics of the problem, given by $\eta_{m}(t)$, takes into account only
the dissipative rates $\Gamma_{mn}$ together with the free evolution
Hamiltonian $\mathcal{H}$ in Eq. (\ref{4}), leaving aside the diffusive
process associated with $\Upsilon_{mn}$. Such a diffusive process, appearing
in the ratio $\varphi(\{\eta_{m}\})/\varphi(\{\eta_{m}(t)\})$ is however,
modified, by the dissipative mechanisms.

\section{Dynamics of the network states: characteristic function,
Glauber-Sudarshan $P$-function, Wigner distribution and density operator}

Starting from two general classes of initial network states, given by mixed
superpositions of coherent and number states, we next analyze the evolution of
such states through the characteristic function, the Glauber-Sudarshan
$P$-function, and the Wigner distribution. We also compute the network density
operator for the case of mixed superposition of Fock states.

\subsection{A mixed superposition of coherent states}

Considering that the initial network state comprehends a mixture of
superpositions of coherent states like $\left\vert \Psi\left(  0\right)
\right\rangle _{\jmath}=%
{\displaystyle\int}
dr_{\jmath}\Lambda\left(  r_{\jmath}\right)  \left\vert \left\{  \beta
_{m}\left(  r_{\jmath}\right)  \right\}  \right\rangle $, the initial density
operator becomes
\begin{equation}
\rho_{S}(0)=\sum_{\jmath}p_{\jmath}%
{\displaystyle\int}
dr_{\jmath}\Lambda\left(  r_{\jmath}\right)
{\displaystyle\int}
ds_{\jmath}\Lambda^{\ast}\left(  s_{\jmath}\right)  \left\vert \left\{
\beta_{m}\left(  r_{\jmath}\right)  \right\}  \right\rangle \left\langle
\left\{  \beta_{m}\left(  s_{\jmath}\right)  \right\}  \right\vert \text{,}
\label{36}%
\end{equation}
where $p_{\jmath}$ is the probability associated with the state $\left\vert
\Psi\left(  0\right)  \right\rangle _{\jmath}$. The parameters $r_{\jmath}$
($s_{\jmath}$) represent a set of variables defining the probability density
function $\Lambda\left(  r_{\jmath}\right)  $, and $\left\vert \left\{
\beta_{m}\left(  r_{\jmath}\right)  \right\}  \right\rangle =\bigotimes
\nolimits_{m}\left\vert \beta_{m}\left(  r_{\jmath}\right)  \right\rangle $
stands for a product of coherent states, where $\left\vert \beta_{m}\left(
r_{\jmath}\right)  \right\rangle $ represents the state associated with the
$m$th network oscillator. In the particular case where $\Lambda\left(
r_{\jmath}\right)  =\sum_{k}\Lambda_{k}\delta\left(  r_{\jmath}-r_{\jmath
}^{\left(  k\right)  }\right)  $, the pure state $\left\vert \Psi\left(
0\right)  \right\rangle _{\jmath}$ becomes the discrete superposition
\begin{equation}
\left\vert \Psi\left(  0\right)  \right\rangle _{\jmath}=\sum_{k}\Lambda
_{k}\left\vert \left\{  \beta_{m}^{k}\right\}  \right\rangle _{\jmath},
\label{37}%
\end{equation}
where we have defined $\left\vert \left\{  \beta_{m}\left(  r_{\jmath
}^{\left(  k\right)  }\right)  \right\}  \right\rangle \equiv\left\vert
\left\{  \beta_{m}^{k}\right\}  \right\rangle _{\jmath}$. Through the
definition of the time-dependent vector elements $K_{m}\left(  r_{\jmath
};t\right)  =\sum_{n}\Theta_{mn}(t)\beta_{n}\left(  r_{\jmath}\right)  $ and
matrix elements
\begin{subequations}
\label{38}%
\begin{align}
\Theta_{mn}(t)  &  =\sum_{\ell}D_{m\ell}\exp\left(  -\Omega_{\ell}t\right)
D_{\ell n}^{-1}\text{,}\label{38a}\\
J_{mn}\left(  t\right)   &  =\Pi_{mn}-\sum_{m^{\prime},n^{\prime}}%
\Pi_{m^{\prime}n^{\prime}}\Theta_{mm^{\prime}}^{\ast}(t)\Theta_{nn^{\prime}%
}(t)\text{,} \label{38c}%
\end{align}
we verify, after a rather lengthy calculation, that the evolution of the
initial network state (\ref{36}) can be described either through the
characteristic function
\end{subequations}
\begin{align}
\chi(\{\eta_{m}\},t)  &  =\sum_{\jmath}p_{\jmath}%
{\displaystyle\int}
ds_{\jmath}\Lambda\left(  s_{\jmath}\right)
{\displaystyle\int}
dr\Lambda^{\ast}\left(  r_{\jmath}\right)  \left\langle \left\{  \beta
_{m}\left(  r_{\jmath}\right)  \right\}  \right.  \left\vert \left\{
\beta_{m}\left(  s_{\jmath}\right)  \right\}  \right\rangle \nonumber\\
&  \times\exp\left\{  \sum_{m}\left[  \eta_{m}K_{m}^{\ast}\left(  r_{\jmath
};t\right)  -\eta_{m}^{\ast}K_{m}\left(  s_{\jmath};t\right)  \right]
-\frac{1}{2}\sum_{m,n}\eta_{m}J_{mn}\left(  t\right)  \eta_{n}^{\ast}\right\}
\text{,} \label{39}%
\end{align}
either by the Glauber-Sudarshan $P$-function%
\begin{align}
P(\{\xi_{m}\},t)  &  =\frac{\left(  2/\pi\right)  ^{N}}{\det\mathbf{J}}%
\sum_{\jmath}p_{\jmath}%
{\displaystyle\int}
ds_{\jmath}\Lambda\left(  s_{\jmath}\right)
{\displaystyle\int}
dr_{\jmath}\Lambda^{\ast}\left(  r_{\jmath}\right)  \left\langle \left\{
\beta_{m}\left(  r_{\jmath}\right)  \right\}  \right.  \left\vert \left\{
\beta_{m}\left(  s_{\jmath}\right)  \right\}  \right\rangle \nonumber\\
&  \times\exp\left\{  -2\sum_{m,n}J_{mn}^{-1}\left(  t\right)  \left[  \xi
_{m}-K_{m}\left(  s_{\jmath};t\right)  \right]  \left[  \xi_{n}-K_{n}\left(
r_{\jmath};t\right)  \right]  ^{\ast}\right\}  \text{,} \label{40}%
\end{align}
or even by the Wigner distribution function
\begin{align}
W(\{\xi_{m}\},t)  &  =\frac{\left(  2/\pi\right)  ^{N}}{\det\mathbf{\tilde{J}%
}}\sum_{\jmath}p_{\jmath}%
{\displaystyle\int}
ds_{\jmath}\Lambda\left(  s_{\jmath}\right)
{\displaystyle\int}
dr_{\jmath}\Lambda^{\ast}\left(  r_{\jmath}\right)  \left\langle \left\{
\beta_{m}\left(  r_{\jmath}\right)  \right\}  \right.  \left\vert \left\{
\beta_{m}\left(  s_{\jmath}\right)  \right\}  \right\rangle \nonumber\\
&  \times\exp\left\{  -2\sum_{m,n}\tilde{J}_{mn}^{-1}\left(  t\right)  \left[
\xi_{m}-K_{m}\left(  s_{\jmath};t\right)  \right]  \left[  \xi_{n}%
-K_{n}\left(  r_{\jmath};t\right)  \right]  ^{\ast}\right\}  \text{.}
\label{41}%
\end{align}

Note that the difference between the Glauber-Sudarshan $P$-function and the
Wigner distribution comes from the time-dependent function associated with the
width of the their Gaussian function. Consequently, the Wigner function can be
obtained from the Glauber-Sudarshan $P$-function through the substitution
$\mathbf{J}\rightarrow\mathbf{\tilde{J}}=\mathbf{J}+\mathbf{I}$. Whereas the
Glauber-Sudarshan $P$-function diverges when there is no diffusion process
such that $\mathbf{J=0}$ (with all the reservoirs at $0$ K), the width of the
Wigner function presents an additional term $\mathbf{I}$ inhibiting any singularity.

For the case of $0$ K reservoirs \cite{MickelGeral}, the density operator of
the network, to be used below, is given by%
\begin{align}
\rho_{S}(t)  &  =\sum_{\jmath}p_{\jmath}%
{\displaystyle\int}
ds_{\jmath}\Lambda\left(  s_{\jmath}\right)
{\displaystyle\int}
dr_{\jmath}\Lambda^{\ast}\left(  r_{\jmath}\right) \nonumber\\
&  \times\frac{\left\langle \left\{  \beta_{m}\left(  r_{\jmath}\right)
\right\}  \right.  \left\vert \left\{  \beta_{m}\left(  s_{\jmath}\right)
\right\}  \right\rangle }{\left\langle \left\{  K_{m}\left(  r_{\jmath
};t\right)  \right\}  \left\vert \left\{  K_{m}\left(  s_{\jmath};t\right)
\right\}  \right.  \right\rangle }\left\vert \left\{  K_{m}\left(  s_{\jmath
};t\right)  \right\}  \right\rangle \left\langle \left\{  K_{m}\left(
r_{\jmath};t\right)  \right\}  \right\vert \text{.} \label{41.5}%
\end{align}

\subsection{A mixed superposition of Fock states}

We now assume the initial network state to be a mixture of superposition of
Fock states $\left\vert \Phi\left(  0\right)  \right\rangle _{\jmath}=%
{\displaystyle\sum\limits_{x_{1},\ldots,x_{N}}}
C_{x_{1},\ldots,x_{N}}^{\left(  \jmath\right)  }\left\vert x_{1},\ldots
,x_{N}\right\rangle $, where the parameter $x_{m}$ indicates the number of
photons in the $m$th oscillator while the coefficient $C_{x_{1},\ldots,x_{N}%
}^{\left(  \jmath\right)  }$ represents the probability amplitude associated
with each state $\left\vert x_{1},\ldots,x_{N}\right\rangle \equiv\left\vert
\left\{  x_{m}\right\}  \right\rangle $ composing the whole superposition. The
initial density operator is thus given by
\begin{equation}
\rho_{S}(0)=\sum_{\jmath}p_{\jmath}%
{\displaystyle\sum\limits_{\left\{  x_{m}\right\}  }}
{\displaystyle\sum\limits_{\left\{  y_{m}\right\}  }}
\left(  C_{\left\{  y_{m}\right\}  }^{\left(  \jmath\right)  }\right)  ^{\ast
}C_{\left\{  x_{m}\right\}  }^{\left(  \jmath\right)  }\left\vert \left\{
x_{m}\right\}  \right\rangle \left\langle \left\{  y_{m}\right\}  \right\vert
\text{,} \label{42}%
\end{equation}
where $p_{\jmath}$ is the probability associated with the state $\left\vert
\Phi\left(  0\right)  \right\rangle _{\jmath}$. Since the Fock state
$\left\vert x_{m}\right\rangle $, of the $m$th oscillator, can be expanded as
a superposition of coherent states of the form
\begin{equation}
\left\vert x_{m}\right\rangle =\mathcal{N}_{m}\int_{0}^{2\pi}d\theta
_{m}\operatorname*{e}\nolimits^{-ix_{m}\theta_{m}}\left\vert \beta
_{m}\operatorname*{e}\nolimits^{i\theta_{m}}\right\rangle , \label{42.1}%
\end{equation}
it is easy to note that the initial state (\ref{42}) can be obtained by Eq.
(\ref{36}), identifying
\begin{subequations}
\label{42.5}%
\begin{align}
\Lambda\left(  r_{\jmath}\right)   &  \rightarrow\Lambda\left(  \theta
_{\jmath}\right)  =%
{\displaystyle\sum\limits_{\left\{  x_{m}\right\}  }}
C_{\left\{  x_{m}\right\}  }^{\left(  \jmath\right)  }\prod\limits_{m}%
\mathcal{N}_{m}\operatorname*{e}\nolimits^{-ix_{m}\theta_{m}}\text{,}%
\label{42.5a}\\%
{\displaystyle\int}
dr_{\jmath}  &  \rightarrow\int_{0}^{2\pi}d\theta_{m}\text{,}\label{42.5b}\\
\beta_{m}\left(  r_{\jmath}\right)   &  \rightarrow\beta_{m}\operatorname*{e}%
\nolimits^{i\theta_{m}}\text{,} \label{42.5c}%
\end{align}
such that we can use the results of the previous subsection to obtain the
characteristic function, the Glauber-Sudarshan $P$-function and Wigner
distribution for a mixed superposition of pure Fock states. Alternatively,
such functions may be directly computed from the initial state (\ref{42}).
Their expressions are presented in Appendix B, where the density operator for
a mixed superposition of pure Fock states is also presented.

\section{Time evolved diffusion coefficients}

To analyze the diffusion mechanism, due to the finite temperature of the
reservoirs, we start by computing the Wigner distribution associated with the
normal-mode oscillators. As depicted in Fig. 2, these oscillators, described
by Hamiltonian $H=H_{0}+V$ [Eqs. (\ref{6a}) and (\ref{6b})], do not interact
with each other, but they do interact with all the reservoirs. We, thus,
rewrite the Wigner distribution (\ref{41}) in a new coordinate frame
$\{\tilde{\xi}_{m}\}$, obtained through the diagonalization of matrix
$\mathbf{\tilde{J}}\left(  t\right)  $. This new framework follows from the
rotation%
\end{subequations}
\begin{equation}
\mathbf{\tilde{\xi}=\xi\bullet U}\left(  t\right)  \text{;\qquad
}\mathbf{\tilde{\xi}^{\ast}=U}^{\dag}\left(  t\right)  \bullet\mathbf{\xi
^{\ast}}\text{,} \label{53}%
\end{equation}
where the unitary operation $\mathbf{U}\left(  t\right)  $ satisfies
$\mathbf{U}^{\dag}\left(  t\right)  \bullet\mathbf{\tilde{J}}\left(  t\right)
\bullet\mathbf{U}\left(  t\right)  =\mathfrak{D}\left(  t\right)  $. From this
matrix relation, we obtain the evolved diffusion coefficients%
\begin{equation}
\mathcal{D}_{m}\left(  t\right)  =\sum_{n,n^{\prime}}U_{mn}^{\dag}\left(
t\right)  \tilde{J}_{nn^{\prime}}\left(  t\right)  U_{n^{\prime}m}\left(
t\right)  \label{56}%
\end{equation}
as the elements of diagonal matrix $\mathfrak{D}\left(  t\right)  $. In this
framework, the rotated Wigner distribution, written as
\begin{equation}
W(\{\tilde{\xi}_{m}\},t)=\sum_{\jmath}p_{\jmath}%
{\displaystyle\int}
ds_{\jmath}%
{\displaystyle\int}
dr_{\jmath}W(\{\tilde{\xi}_{m}\};r_{\jmath},s_{\jmath},t), \label{54}%
\end{equation}
are composed by diagonal ($r_{\jmath}=s_{\jmath}$) and off-diagonal
($r_{\jmath}\neq s_{\jmath}$) elements defined by
\begin{align}
W(\{\tilde{\xi}_{m}\};r_{\jmath},s_{\jmath},t)  &  =\frac{\left(
2/\pi\right)  ^{N}}{\det\mathbf{\tilde{J}}}\Lambda^{\ast}\left(  r_{\jmath
}\right)  \Lambda\left(  s_{\jmath}\right)  \left\langle \left\{  \beta
_{m}\left(  r_{\jmath}\right)  \right\}  \right.  \left\vert \left\{
\beta_{m}\left(  s_{\jmath}\right)  \right\}  \right\rangle \nonumber\\
&  \times\exp\left\{  -\sum_{m}\frac{2}{\mathcal{D}_{m}\left(  t\right)
}\left[  \tilde{\xi}_{m}-\tilde{K}_{m}\left(  s_{\jmath};t\right)  \right]
\left[  \tilde{\xi}_{m}-\tilde{K}_{m}\left(  r_{\jmath};t\right)  \right]
^{\ast}\right\}  \text{,} \label{55}%
\end{align}
where $\mathbf{\tilde{K}}\left(  r_{\jmath};t\right)  =\mathbf{K}\left(
r_{\jmath};t\right)  \bullet\mathbf{U}\left(  t\right)  $ and $\mathbf{\tilde
{K}^{\ast}}\left(  r_{\jmath};t\right)  =\mathbf{U}^{\dag}\left(  t\right)
\bullet\mathbf{K^{\ast}}\left(  r_{\jmath};t\right)  $. The vector
$\mathbf{\tilde{K}}\left(  r_{\jmath};t\right)  $ gives the excitation
intensity of the $m$th normal-mode oscillator through $\left\vert \tilde
{K}_{m}\left(  r_{\jmath};t\right)  \right\vert ^{2}$. We stress that the
larger or smaller values of $\mathcal{D}_{m}\left(  t\right)  $ depend on the
network topology (contained within the matrix elements $U_{mn}$), apart from
the regime of coupling strengths between the oscillators (contained within the
matrix elements $\tilde{J}_{mn}\left(  t\right)  $). As a particular example
of this dependence, we consider a degenerate symmetric network, i.e., a
degenerate network of $N$ oscillators, all of them interacting with each
other, where $\left\{  \omega_{m}\right\}  =\omega$, $\left\{  \lambda
_{mn}\right\}  =\lambda,\left\{  \gamma_{m}\right\}  =\gamma,\left\{
\tilde{\gamma}_{m}\right\}  =\tilde{\gamma}$ and $\left\{  \bar{n}%
_{m}\right\}  =\bar{n}$. In this case, in the strong coupling regime, we
obtain the expression
\begin{equation}
\tilde{J}_{mn}\left(  t\right)  =\delta_{mn}+\frac{2\bar{n}}{N}\left(
1-e^{-N\gamma t}\right)  \text{,} \label{57}%
\end{equation}
and the diffusion coefficients%
\begin{equation}
\mathcal{D}_{m}\left(  t\right)  =\left\{
\begin{array}
[c]{cllll}%
\tilde{J}_{mm}\left(  t\right)  -\tilde{J}_{mn}\left(  t\right)  & = & 1 &
\text{for} & m\neq N\text{,}\\
\tilde{J}_{mm}\left(  t\right)  +\left(  N-1\right)  \tilde{J}_{mn}\left(
t\right)  & = & 1+2\bar{n}\left(  1-e^{-N\gamma t}\right)  & \text{for} &
m=N\text{,}%
\end{array}
\right.  \label{58}%
\end{equation}
showing that $\tilde{J}_{mn}\left(  t\right)  \neq0$ can reduce or enhance the
strength of the diffusion coefficients $\mathcal{D}_{m}\left(  t\right)  $
associated with the normal-mode oscillators.

\subsection{Directional and mean diffusion times}

From the above TD diffusion coefficients (\ref{58}) we define the directional
diffusion time
\begin{equation}
\frac{1}{\tau_{diff}^{\left(  m\right)  }}=\left.  \frac{d}{dt}\mathcal{D}%
_{m}\left(  t\right)  \right\vert _{t=0}\text{,} \label{59}%
\end{equation}
displaying a tendency to a significant spread of the peak --- common to all
elements (the diagonal and off-diagonal) of the Wigner function --- associated
with the $m$th normal-mode oscillator. Since each normal-mode oscillator
defines a direction in the coordinate frame $\{\tilde{\xi}_{m}\}$, we are
naturally led to define the mean diffusion time, associated with all the
dimensions of the space, as the average value%
\begin{equation}
\frac{1}{\tau_{diff}}=\frac{1}{N}\sum_{m}\frac{1}{\tau_{diff}^{\left(
m\right)  }}=\frac{1}{N}\left.  \frac{d}{dt}\operatorname*{Tr}\mathfrak{D}%
\left(  t\right)  \right\vert _{t=0}\text{.} \label{60}%
\end{equation}
The average diffusion time becomes useful to compute the decoherence time of
any network state when complemented with the estimated time for a significant
decay of the peaks associated with the interference terms of the Wigner
function ($r_{\jmath}\neq s_{\jmath}$), to be defined below as $\tau_{int}$,

As an illustrative example of the above theory, below we analyze the diffusion
coefficients $\mathcal{D}_{m}\left(  t\right)  $ for the weak and strong
coupling regimes considering the case of a degenerate symmetric network.

\subsubsection{The weak coupling regime}

In the weak coupling regime, the matrix $\mathbf{\tilde{J}}\left(  t\right)
$, already in a diagonal form, is defined by the elements $\tilde{J}%
_{mn}\left(  t\right)  =\left[  1+2\bar{n}\left(  1-e^{-\gamma t}\right)
\right]  \delta_{mn}$, such that $\mathbf{U=1}$. In this regime, all the
diffusion coefficients equal to%
\begin{equation}
\mathcal{D}_{m}\left(  t\right)  =\mathcal{D}\left(  t\right)  =1+2\bar
{n}\left(  1-e^{-\gamma t}\right)  . \label{61}%
\end{equation}
The average diffusion time becomes
\begin{equation}
\tau_{diff}=\frac{1}{2\bar{n}\gamma}\text{,} \label{62}%
\end{equation}
showing, as expected, that the larger the temperature, the smaller the time
required for a significant diffusion rate. In this case, the coefficients
$\mathcal{D}_{m}\left(  t\right)  $ are mode independent and assume a common
value, such that the spreads of the peaks associated with the diagonal terms
of the Wigner function occurs homogeneously in all directions.

\subsubsection{The strong coupling regime}

In the strong coupling regime, the elements of matrix $\mathbf{\tilde{J}%
}\left(  t\right)  $ are given by Eq. (\ref{57}) and the diffusion
coefficients by Eq. (\ref{58}), showing that only the $N$th normal-mode
oscillator undergoes the diffusion process. For all the normal-mode
oscillators but the $N$th, the diffusion coefficients $\mathcal{D}_{m}\left(
t\right)  $ are counterbalanced by the diffusion rates $\tilde{J}_{mm}\left(
t\right)  $ and $\tilde{J}_{mn}\left(  t\right)  $ coming from the direct- and
indirect-decay channels, respectively. The diffusion coefficients in this
regime lead to the same mean diffusion time as that in Eq. (\ref{62}), showing
that the average diffusion effect comes entirely from the temperatures of the
reservoirs. As to be demonstrated in the next section, this interesting result
is not limited to the degenerate symmetric topology.

\subsection{Diffusion and topology}

Starting from Eq. (\ref{60}) and noting that $\operatorname*{Tr}%
\mathfrak{D}\left(  t\right)  =\operatorname*{Tr}\mathbf{\tilde{J}}\left(
t\right)  $ ($\mathbf{\tilde{J}}\left(  t\right)  =\mathbf{J}\left(  t\right)
+\mathbf{I}$), with the elements of matrix $\mathbf{J}\left(  t\right)  $
given by Eq. (\ref{38c}), we obtain the general expression%
\begin{equation}
\tau_{diff}^{-1}=\frac{2}{N}\operatorname{Tr}\mathbf{\Upsilon}\text{,}
\label{64}%
\end{equation}
applicable to whatever the network topology and the strength coupling regime
between the oscillators, where%
\begin{equation}
\operatorname{Tr}\mathbf{\Upsilon}=N\sum_{m,n}\tilde{\gamma}_{m}(\varpi
_{n})\bar{n}_{m}(\varpi_{n})C_{nm}C_{mn}^{-1}\text{.} \label{65}%
\end{equation}
We note that the information regarding the topology of the network is
contained only in the product $C_{mn}^{-1}C_{nm}$ which acts as a normalized
distribution function ($\sum_{m}C_{nm}C_{mn}^{-1}=1$) when computing the
average value of the diffusion rate given by Eq. (\ref{64}).

We identify two general situations where, as in the case of a degenerate
symmetric network, the diffusion mechanism becomes independent of the topology
of the network. The first situation occurs $i)$ when identical reservoirs are
assumed, such that $\tilde{\gamma}_{m}(\varpi_{n})\bar{n}_{m}(\varpi
_{n})=\tilde{\gamma}(\varpi_{n})\bar{n}(\varpi_{n})$ and, consequently,
$\operatorname{Tr}\mathbf{\Upsilon}=N\sum_{n}\tilde{\gamma}(\varpi_{n})\bar
{n}(\varpi_{n})$, making the mean diffusion%
\begin{equation}
\tau_{diff}^{-1}=2\sum_{m}\tilde{\gamma}(\varpi_{m})\bar{n}(\varpi
_{m})\text{,} \label{66}%
\end{equation}
independent of the network topology. The second situation $ii)$ arises from
the assumptions of Markovian white noise reservoirs and low-temperature
regime, where the normal-mode frequencies satisfy the relation $\hbar
\varpi_{m}\gg k_{B}T$, $k_{B}$ being the Boltzmann constant. In this case we
obtain $\tilde{\gamma}_{m}(\varpi_{n})\bar{n}_{m}(\varpi_{n})\approx
\tilde{\gamma}_{m}\bar{n}_{m}$, such that $\operatorname{Tr}\mathbf{\Upsilon
}=N\sum_{m}\tilde{\gamma}_{m}\bar{n}_{m}$, $\bar{n}_{m}$ being computed around
the average value of the normal-mode frequencies. The mean diffusion time,
independent of the network topology, becomes
\begin{equation}
\tau_{diff}^{-1}=2\sum_{m}\tilde{\gamma}_{m}\bar{n}_{m}. \label{67}%
\end{equation}

Both situation $i)$ and $ii)$ were considered in order to demonstrate that the
mean diffusion time for both, weak and strong coupling regimes, is the same
when considering a degenerate symmetric network. For any other situation,
apart from $i)$ and $ii)$, the average diffusion rate becomes dependent on the
network topology, apart from the reservoirs temperatures.

\section{Collective decoherence rates}

Since an analysis of decoherence through the density operator of the network
is hard to derive when temperature effects are present, it becomes appropriate
to use the Wigner distribution function of the system, instead of the density
operator, to estimate the decoherence time of a family of superposition states
which are particular cases of the general state given by Eq. (\ref{36}). This
family of states is given by
\begin{equation}
\left\vert \psi_{1,\ldots,N}\left(  0\right)  \right\rangle =\mathcal{N}_{\pm
}\left(  \left\vert \underset{R}{\underbrace{\alpha,\ldots,\alpha}}%
,\underset{S}{\underbrace{-\alpha,\ldots,-\alpha}},\underset{N-R-S}%
{\underbrace{\beta,\ldots,\beta}}\right\rangle \pm\left\vert \underset
{R}{\underbrace{-\alpha,\ldots,-\alpha}}\underset{S}{,\underbrace
{\alpha,\ldots,\alpha}},\underset{N-R-S}{\underbrace{\beta,\ldots,\beta}%
}\right\rangle \right)  \mathrm{,} \label{68}%
\end{equation}
where $R$ ($S$) indicates the number of oscillators in the coherent state
$\alpha$ ($-\alpha$) in the first term of the superposition and $-\alpha$
($\alpha$) in the second term of the superposition. The remaining $N-R-S$
oscillators are in the coherent state $\beta$. We again stress that we are
considering a degenerate symmetric network where all the oscillators are
indistinguishable. Therefore, swapping the states of any two oscillators $m$
and $n$, we obtain a state which is completely equivalent to Eq. (\ref{68}).
We also note that when $R=1$ and $S=0$, we obtain from (\ref{68}) the
superposition
\begin{equation}
\left\vert \tilde{\psi}_{1,\ldots,N}\left(  0\right)  \right\rangle
=\mathcal{N}_{\pm}\left(  \left\vert \alpha\right\rangle \pm\left\vert
-\alpha\right\rangle \right)  _{1}\otimes\left\vert \left\{  \beta_{\ell
}\right\}  \right\rangle , \label{69}%
\end{equation}
where a \textquotedblleft Schr\"{o}dinger cat\textquotedblright-like state is
prepared in oscillator $1$ while all the remaining oscillators are prepared in
the coherent states $\beta$.

We start our calculation noting that for a pure two-level state $\left\vert
\Psi\right\rangle =a\left\vert +\right\rangle +b\left\vert -\right\rangle $,
whose density matrix is given by $\rho=a^{\ast}a\left\vert +\right\rangle
\left\langle +\right\vert +b^{\ast}b\left\vert -\right\rangle \left\langle
-\right\vert +a^{\ast}b\left\vert -\right\rangle \left\langle +\right\vert
+ab^{\ast}\left\vert +\right\rangle \left\langle -\right\vert $, the ratio of
the products between the diagonal and off-diagonal elements equals unity,
i.e., $\left(  a^{\ast}b\right)  \left(  ab^{\ast}\right)  /\left(  a^{\ast
}a\right)  \left(  b^{\ast}b\right)  =1$. For an open system described by a
mixed density matrix, however, this ratio decrease from unity. Bearing this in
mind, we rewrite the Wigner function (\ref{54}), to the discrete case where
$\Lambda\left(  r_{\jmath}\right)  =\sum_{k}\Lambda_{k}\delta\left(
r_{\jmath}-r_{\jmath}^{\left(  k\right)  }\right)  $, in a form%
\begin{equation}
W(\{\tilde{\xi}_{m}\},t)=\sum_{r,s=1}^{2}W_{r,s}(\{\tilde{\xi}_{m}\},t),
\label{70}%
\end{equation}
with its diagonal ($r=s$) and off-diagonal ($r\neq s$)\ elements given by%
\begin{align}
W_{r,s}(\{\tilde{\xi}_{m}\},t)  &  =\frac{\left(  2/\pi\right)  ^{N}}%
{\det\mathbf{\tilde{J}}}\Lambda_{r}^{\ast}\Lambda_{s}\left\langle \left\{
\beta_{m}^{r}\right\}  \right.  \left\vert \left\{  \beta_{m}^{s}\right\}
\right\rangle \nonumber\\
&  \times\prod\limits_{m}\exp\left\{  -\frac{2}{\mathcal{D}_{m}\left(
t\right)  }\left[  \tilde{\xi}_{m}-\tilde{K}_{m}^{s}\left(  t\right)  \right]
\left[  \tilde{\xi}_{m}-\tilde{K}_{m}^{r}\left(  t\right)  \right]  ^{\ast
}\right\}  , \label{71}%
\end{align}
where $r$ and $s$ (running from $1$ to $2$) label the product states composing
the superposition (\ref{68}).

Now, through the diagonal and off-diagonal elements of the Wigner function, we
define the ratio%
\begin{align}
\Xi_{rs}(t)  &  =\frac{W_{r,r}(\{\tilde{\xi}_{m}\},t)W_{s,s}(\{\tilde{\xi}%
_{m}\},t)}{W_{r,s}(\{\tilde{\xi}_{m}\},t)W_{s,r}(\{\tilde{\xi}_{m}%
\},t)}\nonumber\\
&  =\exp\left[  \sum_{m}\left(  \left\vert \beta_{m}^{s}-\beta_{m}%
^{r}\right\vert ^{2}-\frac{2}{\mathcal{D}_{m}\left(  t\right)  }\left\vert
\sum_{m^{\prime},n}U_{mm^{\prime}}\left(  t\right)  \Theta_{m^{\prime}%
n}\left(  t\right)  \left(  \beta_{n}^{r}-\beta_{n}^{s}\right)  \right\vert
^{2}\right)  \right]  \text{.} \label{72}%
\end{align}
which turns to be independent on the variables $\{\tilde{\xi}_{m}\}$ of the
Wigner function, as desired. Moreover, for $t=0$, such that $\Theta
_{mn}(0)=\delta_{mn}$ and $\mathcal{D}_{m}\left(  0\right)  =1$, we obtain
$\Xi_{rs}(0)=\exp\left(  -\sum_{m}\left\vert \beta_{m}^{s}-\beta_{m}%
^{r}\right\vert ^{2}\right)  $. In analogy with the above observation
concerning the ratio of the products between the diagonal and off-diagonal
elements of a pure or mixed density matrix, the above defined ratio $\Xi
_{rs}(t)$ offers a measure of the decoherence rate which follows from the
function
\begin{align}
\wp_{rs}\left(  t\right)   &  \equiv\frac{\Xi_{rs}(0)}{\Xi_{rs}(t)}\nonumber\\
&  =\exp\left[  -2\sum_{m}\left(  \left\vert \beta_{m}^{s}-\beta_{m}%
^{r}\right\vert ^{2}-\frac{1}{\mathcal{D}_{m}\left(  t\right)  }\left\vert
\sum_{m^{\prime},n}U_{mm^{\prime}}\left(  t\right)  \Theta_{m^{\prime}%
n}\left(  t\right)  \left(  \beta_{n}^{r}-\beta_{n}^{s}\right)  \right\vert
^{2}\right)  \right]  \text{,} \label{73}%
\end{align}
which equals unity for $t=0$. The above deduction of the decay function
(\ref{73}) can also be developed for the general case of an initial continuous
superposition state, instead of a discrete one.

\subsection{The equivalence between the decays of the interference terms of
both the Wigner Function and the density operator: reservoirs at absolute
zero}

This subsection is devoted to demonstrate that the measure of the decoherence
rate offered by Eq. (\ref{73}) is equivalent to the one coming from the
interference terms of the density operator, which is commonly used for the
case of 0$K$ reservoirs. In fact, for reservoirs at 0$K$, where $\mathcal{D}%
_{m}\left(  t\right)  =1$, it is simple to verify that Eq. (\ref{73}) reduces
to%
\begin{align}
\wp_{rs}\left(  t\right)   &  =\exp\left[  -2\sum_{m}\left(  \left\vert
\beta_{m}^{s}-\beta_{m}^{r}\right\vert ^{2}-\left\vert \sum_{n}\Theta
_{mn}(t)\left(  \beta_{n}^{s}-\beta_{n}^{r}\right)  \right\vert ^{2}\right)
\right] \nonumber\\
&  =\left\vert \frac{\left\langle \left\{  \beta_{m}^{r}\right\}  \left\vert
\left\{  \beta_{m}^{s}\right\}  \right.  \right\rangle }{\left\langle \left\{
K_{m}^{r}\left(  t\right)  \right\}  \left\vert \left\{  K_{m}^{s}\left(
t\right)  \right\}  \right.  \right\rangle }\right\vert ^{4}\text{,}
\label{73.5}%
\end{align}
where the coefficients $\left\langle \left\{  \beta_{m}^{r}\right\}
\left\vert \left\{  \beta_{m}^{s}\right\}  \right.  \right\rangle
/\left\langle \left\{  K_{m}^{r}\left(  t\right)  \right\}  \left\vert
\left\{  K_{m}^{s}\left(  t\right)  \right\}  \right.  \right\rangle $ are
those coming from the interference terms of density operator (\ref{41.5}),
when considering a discrete case. Therefore, considering that decoherence
times are usually estimated through the relation $\left\langle \left\{
\beta_{m}^{r}\right\}  \left\vert \left\{  \beta_{m}^{s}\right\}  \right.
\right\rangle /\left\langle \left\{  K_{m}^{r}\left(  \tau_{D}\right)
\right\}  \left\vert \left\{  K_{m}^{s}\left(  \tau_{D}\right)  \right\}
\right.  \right\rangle =e^{-1}$, for the case of reservoirs at 0$K$, we obtain
from Eq. (\ref{73.5}) the equivalent relation $\wp_{rs}\left(  \tau
_{D}\right)  =e^{-4}$, which gives the estimative of the decoherence time
through the Wigner function.

\subsection{Decay time of the interference terms}

Now we are able to define the time $\tau_{int}$ for a significant decay of the
peaks associated with the interference terms of the Wigner function
($r_{\jmath}\neq s_{\jmath}$). This is done, by generalizing the relation
$\wp_{rs}\left(  \tau_{D}\right)  =e^{-4}$, for the case of reservoirs at
finite temperatures, to the equality
\begin{equation}
\wp_{rs}\left(  \tau_{int}\right)  =\exp\left[  -\left.  4N\right/  \sum
_{m}\mathcal{D}_{m}\left(  \tau_{int}\right)  \right]  \text{,}%
\end{equation}
that corresponds to measure the decay of the interference terms of the Wigner
function by deducting their spreadings, common to all the diagonal and
off-diagonal elements, as we can see in Eq.(\ref{71}). In other words, it is
similar to analyze the decay of the interference terms in a frame where the
diagonal terms are frozen.

\subsection{Decoherence time}

Finally, to define a decoherence time $\tau_{D}$, which take into account both
the diffusion and decay of the interference terms, we must consider both the
above defined times: the mean diffusion time $\tau_{diff}$ and the decay time
of the interference terms of the Wigner functions $\tau_{int}$. We thus define
the relation
\begin{equation}
\frac{1}{\tau_{D}}=\frac{1}{\tau_{diff}}+\frac{1}{\tau_{int}}\text{,}%
\label{74}%
\end{equation}
where $\tau_{diff}^{-1}$ only becomes relevant for particular initial states
whose interference terms of the Wigner function are null, as occur, for
example in the case $N=1$, to the coherent state $\left\vert \alpha
\right\rangle $, or when the excitation of the components of a superposition
state is significantly smaller than unity. This will become clear in the
example to be analyzed below for the degenerate symmetric network. In the
first case, it is well-known that a coherent state remains as such, even under
a dissipative process, when considering a reservoir at $0$ K. However, when
considering a reservoir at finite temperature, the decoherence time of a
coherent state $\left\vert \alpha\right\rangle $ can be estimated through our
defined Eqs. (\ref{64}) and (\ref{65}).

\subsubsection{The weak coupling regime}

The Wigner function associated with the pure state (\ref{68}) in the weak
coupling regime, is obtained from Eq. (\ref{71}) with $\mathbf{U=1}$ and
$\mathcal{D}_{m}\left(  t\right)  =\mathcal{D}\left(  t\right)  =1+2\bar
{n}\left(  1-e^{-\gamma t}\right)  $. Our defined decay function (\ref{73})
thus becomes%
\begin{equation}
\wp_{rs}\left(  t\right)  \equiv\exp\left[  -8\mathcal{D}^{-1}\left(
t\right)  \left\vert \alpha\right\vert ^{2}\left(  R+S\right)  \left(
1+2\bar{n}\right)  \left(  1-e^{-\gamma t}\right)  \right]  \text{.}
\label{76}%
\end{equation}
We estimate the decoherence time $\tau_{D}$ of the family of states (\ref{68})
through the equality $\wp_{rs}\left(  \tau_{int}\right)  =\exp\left[
-4\mathcal{D}^{-1}\left(  \tau_{int}\right)  \right]  $. The obtained result
for the decay time and so for the decoherence time reads%
\begin{equation}
\tau_{D}\approx\tau_{int}=\frac{1}{2\left\vert \alpha\right\vert ^{2}\gamma
}\frac{1}{\left(  R+S\right)  \left(  1+2\bar{n}\right)  }\text{,} \label{77}%
\end{equation}
which recover the results in Ref. \cite{MickelDFS} for 0$K$ reservoirs
($\bar{n}=0$). In Eq. (\ref{77}) we have disregarded the mean diffusion time
$\tau_{diff}^{-1}=2\bar{n}\gamma$ since we assumed that the excitation
$\left(  R+S\right)  \left\vert \alpha\right\vert ^{2}$ is significantly
larger than unity. Note that in the case where $R=N$ ($S=0$) or $S=N$ ($R=0$),
given the initial entangled state $\left\vert \hat{\psi}_{1,\ldots,N}\left(
0\right)  \right\rangle =\mathcal{N}_{\pm}\left(  \left\vert \alpha
,\ldots,\alpha,\right\rangle \pm\left\vert -\alpha,\ldots,-\alpha\right\rangle
\right)  $, the decoherence time decreases as the number of network
oscillators increases.

For the case of the \textquotedblleft Schr\"{o}dinger cat\textquotedblright%
-like state in Eq. (\ref{69}), we obtain the result
\begin{equation}
\tau_{D}\approx\tau_{int}=\frac{1}{2\left\vert \alpha\right\vert ^{2}\gamma
}\frac{1}{\left(  1+2\bar{n}\right)  }\text{.} \label{Schr}%
\end{equation}
Summarizing, the temperature effect decreases the decoherence time when the
weak coupling regime is considered.

\subsubsection{The strong coupling regime}

From the Wigner function associated with the state $\left\vert \hat{\psi
}_{1,\ldots,N}\left(  0\right)  \right\rangle $, derived from Eq. (\ref{71})
and using the coefficients (\ref{58}), we obtain in the strong coupling regime%
\begin{equation}
\wp_{rs}\left(  t\right)  =\exp\left[  -8\mathcal{D}_{N}^{-1}\left(  t\right)
\left\vert \alpha\right\vert ^{2}N^{2}\left(  1+2\bar{n}\right)  \left(
1-e^{-\gamma Nt}\right)  /N\right]  \text{.} \label{78}%
\end{equation}
The estimated decay time $\tau_{int}$ of the interference terms of the Wigner
functions is established through the inequality $\wp_{rs}\left(  \tau
_{int}\right)  =\exp\left\{  -4N/\left[  N-1+\mathcal{D}_{N}\left(  \tau
_{int}\right)  \right]  \right\}  \leq\exp\left\{  -4\mathcal{D}_{N}%
^{-1}\left(  \tau_{int}\right)  \right\}  $, such that%
\begin{equation}
\tau_{int}\geq\frac{1}{2\left\vert \alpha\right\vert ^{2}\gamma}\frac{1}%
{N^{2}\left(  1+2\bar{n}\right)  }\text{,} \label{80}%
\end{equation}
showing that the interference terms of the Wigner distribution decay at a
fastest rate than in the weak coupling regime. For the \textquotedblleft
Schr\"{o}dinger cat\textquotedblright-like state, Eq. (\ref{69}), we obtain
exactly the result shown in Eq. (\ref{Schr}).

We finally note that, considering only the usual decay of the interference
terms, given by $\wp_{rs}\left(  \tau_{D}\right)  =\exp\left(  -4\right)  $,
the estimation of the decoherence time leads to inconsistent results which
present negative values apart from singularities. For example, for the
\textquotedblleft Schr\"{o}dinger cat\textquotedblright-like state in Eq.
(\ref{69}), in the particular case $N=1$, we obtain
\begin{equation}
\tau_{D}\approx\frac{1}{2\gamma\left[  \left\vert \alpha\right\vert
^{2}\left(  1+2\bar{n}\right)  -\bar{n}\right]  },
\end{equation}
which has a singularities at $\bar{n}=\left\vert \alpha\right\vert
^{2}/\left(  1-2\left\vert \alpha\right\vert ^{2}\right)  $ and becomes
negative for $\left\vert \alpha\right\vert ^{2}\left(  1+2\bar{n}\right)  <$
$\bar{n}$. Therefore, the procedure adopted in Eq. (\ref{74}) to estimate the
decoherence time by separating both effects of diffusion and decay of the
Wigner function interference terms, is in fact more sound than the cruder
approach where only the interference effects present in the decay function
(\ref{73}) are considered. Another example refers to the decoherence of a
coherent state $\left\vert \alpha\right\rangle $, where the result $\tau
_{D}\approx1/2\gamma\bar{n}$ computed though the our technique, account
exactly for the diffusion effect, apart from the decay rate $\gamma$, as
expected. The usual procedure fails to give such an account.

As mentioned above, the analysis of the emergence of DFSs with the reservoirs
at finite temperature is addressed in another work \cite{Mickel4}, where both,
collective effects of damping and diffusion, are managed together with the
network topology to build up desired DFSs.

\section{Computing the entropy and the entanglement degree through the Wigner
function}

The computation of the density operator of the network for the case of
reservoirs at finite temperatures becomes a difficult task for the majority of
the initial network states. Therefore, similarly to our procedure to the
analysis of decoherence, we next compute the entropy of the network using the
Wigner functions as given by%
\begin{align}
S &  =1-\operatorname{Tr}\rho_{S}^{2}=1-\pi^{N}\int_{-\infty}^{\infty}%
d^{2}\left\{  \xi_{m}\right\}  W^{2}(\{\xi_{m}\},t)\nonumber\\
&  =1-\pi^{N}\int_{-\infty}^{\infty}d^{2}\left\{  \tilde{\xi}_{m}\right\}
W^{2}(\{\tilde{\xi}_{m}\},t)\text{,}%
\end{align}
where the factor $\pi^{N}$ was introduced to produce a null lower bound for
the entropy. Using the integral result
\begin{equation}
\frac{1}{\pi}\int d^{2}\eta_{m}\exp\left(  a_{m}\eta_{m}^{\ast}-b_{m}\eta
_{m}-c_{m}\eta_{m}^{\ast}\eta_{m}\right)  =\frac{1}{c_{m}}\exp\left(
-\frac{a_{m}b_{m}}{c_{m}}\right)  \text{,}%
\end{equation}
and the Wigner function given by Eq. (\ref{41}), or Eq. (\ref{55}), we obtain
the general expression%
\begin{align}
S\left(  t\right)   &  =1-%
{\displaystyle\int}
dr%
{\displaystyle\int}
dr^{\prime}%
{\displaystyle\int}
ds%
{\displaystyle\int}
ds^{\prime}\Lambda^{\ast}\left(  r\right)  \Lambda^{\ast}\left(  r^{\prime
}\right)  \Lambda\left(  s\right)  \Lambda\left(  s^{\prime}\right)
\left\langle \left\{  \beta_{m}\left(  r\right)  \right\}  \right.  \left\vert
\left\{  \beta_{m}\left(  s\right)  \right\}  \right\rangle \nonumber\\
&  \times\left\langle \left\{  \beta_{m}\left(  r^{\prime}\right)  \right\}
\right.  \left\vert \left\{  \beta_{m}\left(  s^{\prime}\right)  \right\}
\right\rangle \mathcal{P}_{rs,r^{\prime}s^{\prime}}\left(  t\right)  ,
\end{align}
which is applicable to any initial network state, where%
\begin{align}
\mathcal{P}_{rs,r^{\prime}s^{\prime}}\left(  t\right)   &  =\frac{1}%
{\det\mathbf{J}}\exp\left\{  -\sum_{m}\left[  \upsilon_{m}\left(  s,s^{\prime
}\right)  \upsilon_{m}^{\ast}\left(  r,r^{\prime}\right)  \right.  \right.
\nonumber\\
&  \left.  \left.  -\frac{1}{\mathfrak{D}_{m}\left(  t\right)  }\left(
\sum_{\ell,n}U_{\ell m}\left(  t\right)  \Theta_{\ell n}\left(  t\right)
\upsilon_{n}\left(  s,s^{\prime}\right)  \right)  \left(  \sum_{\ell,n}U_{\ell
m}\left(  t\right)  \Theta_{\ell n}\left(  t\right)  \upsilon_{n}\left(
r,r^{\prime}\right)  \right)  ^{\ast}\right]  \right\}  \text{,}\label{P}%
\end{align}
and $\upsilon_{m}\left(  r,s\right)  =\beta_{m}\left(  r\right)  -\beta
_{m}\left(  s\right)  .$ For the case where dissipation is absent, i.e.,
$\gamma_{m}(\omega)=\tilde{\gamma}_{m}(\omega)=0$, we verify that
$\mathcal{P}_{rs,r^{\prime}s^{\prime}}\left(  t\right)  =1$ and, consequently,
$S=0$. Oppositely, when $\gamma_{m}(\omega)\neq0$ and $\tilde{\gamma}%
_{m}(\omega)\neq0$, the purity loss follows from the decay of $\mathcal{P}%
_{rs,r^{\prime}s^{\prime}}\left(  t\right)  $ which reduces to the function
$\wp_{rs}\left(  t\right)  ,$Eq. (\ref{73}), that enters in the calculation of
the decoherence time, under the conditions $\det\mathbf{J}=1$ ($T=0$ K$)$,
$r=r^{\prime}$, and $s=s^{\prime}$. As expected, the purity loss mechanism is
intimately related to the decoherence one.

Focusing on the case when  $\gamma_{m}(\omega)=\tilde{\gamma}_{m}(\omega)=0$,
the entanglement degree of a bipartite system, described by a \emph{pure}
density operator $\rho_{AB}$ -- $A$ and $B$ standing for two complementary
sets of network oscillators --, can be computed through the reduced entropy
(concurrence)%
\begin{equation}
\mathcal{C}=1-\operatorname{Tr}_{A}\left[  \operatorname{Tr}_{B}\rho
_{AB}\right]  ^{2}=1-\operatorname{Tr}_{B}\left[  \operatorname{Tr}_{A}%
\rho_{AB}\right]  ^{2}\text{,}%
\end{equation}
which is given, through the joint Wigner function $W\left(  \left\{  \xi
_{A}\right\}  ,\left\{  \xi_{B}\right\}  ,t\right)  $, as%
\begin{align}
\mathcal{C} &  =1-\pi^{N_{A}}\int_{-\infty}^{\infty}d^{2}\left\{  \xi
_{A}\right\}  \left[  \int_{-\infty}^{\infty}d^{2}\xi_{B}W\left(  \left\{
\xi_{A}\right\}  ,\left\{  \xi_{B}\right\}  ,t\right)  \right]  ^{2}%
\nonumber\\
&  =1-\pi^{N_{B}}\int_{-\infty}^{\infty}d^{2}\left\{  \xi_{B}\right\}  \left[
\int_{-\infty}^{\infty}d^{2}\xi_{A}W\left(  \left\{  \xi_{A}\right\}
,\left\{  \xi_{B}\right\}  ,t\right)  \right]  ^{2}\text{,}%
\end{align}
where $N_{A}$ and $N_{B}$ refer to the numbers of oscillators composing the
sets $A$ and $B$, respectively.

When the subsystems $A$ and $B$ are uncorrelated, such that $\rho_{AB}%
=\rho_{A}\otimes\rho_{B}$ , the Wigner function is factorized as
\begin{equation}
W\left(  \left\{  \xi_{A}\right\}  ,\left\{  \xi_{B}\right\}  ,t\right)
=W\left(  \left\{  \xi_{A}\right\}  ,t\right)  W\left(  \left\{  \xi
_{B}\right\}  ,t\right)  \text{,}%
\end{equation}
and, consequently%
\begin{equation}
\mathcal{C}=1-\pi^{N_{B}}\int_{-\infty}^{\infty}d^{2}\left\{  \xi_{B}\right\}
W^{2}\left(  \left\{  \xi_{B}\right\}  ,t\right)  =0\text{,}%
\end{equation}
as expected.

\section{Concluding remarks}

In the present work we have analyzed the effects of temperature in a network
of dissipative quantum harmonic oscillators. Starting from a previous work
where a general treatment of such a bosonic dissipative network was presented
\cite{MickelGeral}, in the case of reservoirs at 0$K$, here we considered
reservoirs at finite temperatures. Through the solution obtained for the
normal-ordered characteristic function, we did compute formal expressions for
the Glauber-Sudarshan $P$-function, the Wigner distribution function, and the
density operator for whichever the initial network state. An important point
to be stressed is the relevance played by the Wigner function in the present
context where the reservoirs are at finite temperature. In fact, it becomes
hard to identify the main features associated with the dynamic of the network
states through the density operator which results to be an intricate
expression. Through the Wigner function, however, the diffusion coefficients
of the normal-mode oscillators are clearly identified as well as the decay of
its interference terms. We also showed how to compute the entropy and the
entanglement degree through the Wigner function.

We demonstrated that the diffusion coefficients $\mathcal{D}_{m}\left(
t\right)  $ associated with the normal-mode oscillators present completely
different behaviors in both weak and strong coupling regimes. In the former
case, where the indirect-decay channels do not take place, the diffusion
coefficients are entirely related to the dissipative processes of the
oscillators to their own reservoirs. In this case the collective damping and
diffusion effects are dismissible. However, in the later case, the diffusion
coefficients $\mathcal{D}_{m}\left(  t\right)  $ are counterbalanced by the
diffusion rates $\tilde{J}_{mm}\left(  t\right)  $ and $\tilde{J}_{mn}\left(
t\right)  $ coming from the direct- and indirect-decay channels, respectively.
In this case, the collective damping and diffusion effects emerges from the
fact that all network oscillators interact with all the reservoirs due to the
strong coupling between each other. In fact, in the strong coupling regime,
the individual oscillators cannot account for the dynamic of the whole
network, which must be described through the collective normal-mode
oscillators. Differently, in the weak coupling regime, the network dynamic
follows directly from those of the individual oscillators.

In sum, we have presented an analysis of the mechanisms for handling the
diffusion coefficients $\mathcal{D}_{m}\left(  t\right)  $ in the strong
coupling regime, by manipulating the diffusion rates $\tilde{J}_{mm}\left(
t\right)  $ and $\tilde{J}_{mn}\left(  t\right)  $ through the nature and the
temperature of the reservoirs, apart from the network topology. Such approach
was explored in Ref. \cite{Mickel4} to demonstrate the possibility of the
emergence of DFSs in a network of dissipative oscillators even with the
reservoirs at finite temperatures.

We have also present a technique to estimate the decoherence time of network
states which separates effects of diffusion from the decay of the interference
terms in the Wigner distribution function. Our technique overcomes the
difficulties that show up with negative values and singularities arising from
the usual definition of the decoherence time based only on the decay of
interference terms. We have computed the decoherence time for some particular
states of the network, leaving for another work \cite{Mickel4} the analysis of
the emergence of DFSs under temperature effects.

\appendix{}

\section{Matrix equation}

The solution of an arbitrary matrix equation of the form $\mathbf{M\bullet
X}+\mathbf{X\bullet N}^{\top}=\mathbf{P}$ (for an unknown $\mathbf{X}$) can be
obtained through the solution of the system%
\begin{equation}
\left[  \left(  \mathbf{I}\otimes\mathbf{M}\right)  +\left(  \mathbf{N}%
\otimes\mathbf{I}\right)  \right]  \bullet\operatorname{vec}\left(
\mathbf{X}\right)  \equiv\operatorname{vec}\left(  \mathbf{P}\right)  \text{,}
\label{89}%
\end{equation}
following from the inverse of $\left[  \left(  \mathbf{I}\otimes
\mathbf{M}\right)  +\left(  \mathbf{N}\otimes\mathbf{I}\right)  \right]  $,
given by
\begin{equation}
\operatorname{vec}\left(  \mathbf{X}\right)  =\left[  \left(  \mathbf{I}%
\otimes\mathbf{M}\right)  +\left(  \mathbf{N}\otimes\mathbf{I}\right)
\right]  ^{-1}\bullet\operatorname{vec}\left(  \mathbf{P}\right)  \text{,}
\label{90}%
\end{equation}
where the notation $\operatorname{vec}\left(  \mathbf{P}\right)  $ was defined
in Eq.(\ref{30}). Before computing the elements of the inverse matrix $\left(
\mathbf{I}\otimes\mathbf{M}\right)  +\left(  \mathbf{N}\otimes\mathbf{I}%
\right)  \equiv\mathbf{Q}$, it is useful to observe some important properties
of $\mathbf{Q}$:

$i)$ The eigenvalues of matrix $\mathbf{Q}$, defined by $\varepsilon_{ij}$,
are obtained through the direct sum of the eigenvalues $\epsilon_{i}$ and
$\tilde{\epsilon}_{i}$ of matrices $\mathbf{M}$ and $\mathbf{N}$, such that%
\begin{equation}
\varepsilon_{ij}=\tilde{\epsilon}_{i}+\epsilon_{j}\text{.} \label{91}%
\end{equation}

$ii)$ The eigenvectors of matrix $\mathbf{Q}$ are obtained through the tensor
product
\begin{equation}
\vartheta_{ij}=\tilde{\nu}^{\left(  i\right)  }\otimes\nu^{\left(  j\right)
}, \label{92}%
\end{equation}
where $\nu^{\left(  i\right)  }$ and $\tilde{\nu}^{\left(  i\right)  }$
describe the eigenvector associated to the eigenvalue $\epsilon_{i}$ and
$\tilde{\epsilon}_{i}$. In fact, knowing the eigenvalues and eigenvectors of
matrices $\mathbf{M}$ and $\mathbf{N}$, we can easily verify that
$\vartheta_{ij}$ defines the desired eigenvector, associated with the
eigenvalue $\varepsilon_{ij}$, since
\begin{align}
\left[  \left(  \mathbf{I}\otimes\mathbf{M}\right)  +\left(  \mathbf{N}%
\otimes\mathbf{I}\right)  \right]  \bullet\left(  \tilde{\nu}^{\left(
i\right)  }\otimes\nu^{\left(  j\right)  }\right)   &  =\tilde{\nu}^{\left(
i\right)  }\otimes\left(  \mathbf{M}\bullet\nu^{\left(  j\right)  }\right)
+\left(  \mathbf{N}\bullet\tilde{\nu}^{\left(  i\right)  }\right)  \otimes
\nu^{\left(  j\right)  }\nonumber\\
&  =\left(  \tilde{\epsilon}_{i}+\epsilon_{j}\right)  \left(  \tilde{\nu
}^{\left(  i\right)  }\otimes\nu^{\left(  j\right)  }\right) \nonumber\\
&  =\varepsilon_{ij}\left(  \tilde{\nu}^{\left(  i\right)  }\otimes
\nu^{\left(  j\right)  }\right)  \text{.} \label{93}%
\end{align}

\section{Alternative expression for the evolution of a mixed superposition of
Fock states}

We verify that the evolution of the initial state (\ref{42}) can be
characterized, using the same definitions (\ref{38a}) and (\ref{38c}), through
the characteristic function%
\begin{align}
&  \chi(\{\eta_{m}\},t)=\sum_{\jmath}%
{\displaystyle\sum\limits_{\left\{  x_{m}\right\}  }}
{\displaystyle\sum\limits_{\left\{  y_{m}\right\}  }}
p_{\jmath}\left(  C_{\left\{  y_{m}\right\}  }^{\left(  \jmath\right)
}\right)  ^{\ast}C_{\left\{  x_{m}\right\}  }^{\left(  \jmath\right)  }%
\exp\left(  -\frac{1}{2}\sum_{m,n}\eta_{m}J_{mn}\left(  t\right)  \eta
_{n}^{\ast}\right) \nonumber\\
&  \times%
{\displaystyle\prod\limits_{\ell}}
\left[  \sum_{j_{\ell}=0}^{x_{\ell}}\frac{\sqrt{y_{\ell}!x_{\ell}!}}{j_{\ell
}!\left(  x_{\ell}-j_{\ell}\right)  !\left(  y_{\ell}-x_{\ell}+j_{\ell
}\right)  !}\left(  \sum_{m}\eta_{m}\Theta_{m\ell}^{\ast}(t)\right)
^{y_{\ell}-x_{\ell}+j_{\ell}}\left(  -\sum_{m}\eta_{m}^{\ast}\Theta_{m\ell
}(t)\right)  ^{j_{\ell}}\right]  , \label{43}%
\end{align}
which leads to the Wigner distribution function given in terms of derivatives
as%
\begin{align}
W(\{\xi_{m}\},t)  &  =\frac{\left(  2/\pi\right)  ^{N}}{\det\mathbf{\tilde{J}%
}}\sum_{\jmath}%
{\displaystyle\sum\limits_{\left\{  x_{m}\right\}  }}
{\displaystyle\sum\limits_{\left\{  y_{m}\right\}  }}
p_{\jmath}\left(  C_{\left\{  y_{m}\right\}  }^{\left(  \jmath\right)
}\right)  ^{\ast}C_{\left\{  x_{m}\right\}  }^{\left(  \jmath\right)
}\nonumber\\
&  \times\left(
{\displaystyle\prod\limits_{\ell}}
\sum_{j_{\ell}=0}^{x_{\ell}}\frac{\sqrt{y_{\ell}!x_{\ell}!}}{j_{\ell}!\left(
x_{\ell}-j_{\ell}\right)  !\left(  y_{\ell}-x_{\ell}+j_{\ell}\right)  !}%
\lim_{\varepsilon_{\ell}\rightarrow0}\frac{\partial^{y_{\ell}-x_{\ell
}+2j_{\ell}}}{\partial\left(  \varepsilon_{\ell}\right)  ^{j_{\ell}}%
\partial\left(  \varepsilon_{\ell}^{\ast}\right)  ^{y_{\ell}-x_{\ell}+j_{\ell
}}}\right) \nonumber\\
&  \times\exp\left[  -2\sum_{m,n}\tilde{J}_{mn}^{-1}\left(  t\right)  \left(
\xi_{m}-\sum_{\ell}\varepsilon_{\ell}\Theta_{m\ell}(t)\right)  \left(  \xi
_{n}^{\ast}-\sum_{\ell}\varepsilon_{\ell}^{\ast}\Theta_{n\ell}^{\ast
}(t)\right)  \right]  . \label{44}%
\end{align}
The above distribution can also be given explicitly in the form
\begin{align}
&  W(\{\xi_{m}\},t)=\frac{\left(  2/\pi\right)  ^{N}}{\det\mathbf{\tilde{J}}%
}\sum_{\jmath}%
{\displaystyle\sum\limits_{\left\{  x_{m}\right\}  }}
{\displaystyle\sum\limits_{\left\{  y_{m}\right\}  }}
p_{\jmath}\left(  C_{\left\{  y_{m}\right\}  }^{\left(  \jmath\right)
}\right)  ^{\ast}C_{\left\{  x_{m}\right\}  }^{\left(  \jmath\right)  }\left(
%
{\displaystyle\prod\limits_{\ell}}
\sum_{q_{\ell}=0}^{x_{\ell}}\frac{\sqrt{y_{\ell}!x_{\ell}!}}{\left(  x_{\ell
}-q_{\ell}\right)  !}\right) \nonumber\\
&  \times\left[
{\displaystyle\prod\limits_{\ell,\ell^{\prime}}}
\sum_{R_{\ell\ell^{\prime}}=0}^{R_{\ell,\ell^{\prime}-1}}\Delta_{\ell^{\prime
}\ell}(\left\{  R_{n,n^{\prime}}\right\}  ;t)\digamma\left(  y_{\ell^{\prime}%
}-x_{\ell^{\prime}}+q_{\ell^{\prime}}-\sum_{i=1}^{\ell-1}\left(
R_{i,\ell^{\prime}-1}-R_{i,\ell^{\prime}}\right)  -\left(  R_{\ell
,\ell^{\prime}-1}-R_{\ell,\ell^{\prime}}\right)  \right)  \right] \nonumber\\
&  \times\left(
{\displaystyle\prod\limits_{\ell}}
\Lambda_{\ell}(\left\{  R_{n,n^{\prime}}\right\}  ,\left\{  \xi_{p}\right\}
;t)\frac{\left[  2\sum_{m,n}\tilde{J}_{mn}^{-1}\left(  t\right)  \Theta
_{m\ell}(t)\xi_{n}^{\ast}\right]  ^{R_{\ell N}}}{R_{\ell N}!}\right)
\exp\left(  -2\sum_{m,n}\xi_{m}\tilde{J}_{mn}^{-1}\xi_{n}^{\ast}\right)  ,
\label{45}%
\end{align}
where we have defined
\begin{subequations}
\label{46}%
\begin{align}
R_{\ell,0}  &  =q_{\ell}\text{,}\label{46a}\\
\Delta_{mn}(\left\{  R_{\ell,\ell^{\prime}}\right\}  ;t)  &  =\frac{1}{\left(
R_{n,m-1}-R_{n,m}\right)  !}\left[  -2\sum_{\ell,\ell^{\prime}}\tilde{J}%
_{\ell\ell^{\prime}}^{-1}\left(  t\right)  \Theta_{\ell n}(t)\Theta
_{\ell^{\prime}m}^{\ast}(t)\right]  ^{R_{n,m-1}-R_{n,m}}\text{,}\label{46b}\\
\Lambda_{m}(\left\{  R_{\ell,\ell^{\prime}}\right\}  ,\left\{  \xi
_{p}\right\}  ;t)  &  =\frac{\left[  2\sum_{\ell,\ell^{\prime}}\tilde{J}%
_{\ell\ell^{\prime}}^{-1}\left(  t\right)  \xi_{\ell}\Theta_{\ell^{\prime}%
m}^{\ast}(t)\right]  ^{y_{m}-x_{m}+q_{m}-\sum_{\ell}\left(  R_{\ell
,m-1}-R_{\ell,m}\right)  }}{\left[  y_{m}-x_{m}+q_{m}-\sum_{\ell}\left(
R_{\ell,m-1}-R_{\ell,m}\right)  \right]  !}\text{,} \label{46c}%
\end{align}
with%
\end{subequations}
\begin{equation}
\digamma\left(  x\right)  =\left\{
\begin{array}
[c]{ccc}%
1 & \text{for} & x\geq0\\
0 & \text{for} & x<0
\end{array}
\right.  \text{.} \label{47}%
\end{equation}

As noted in Section IV, we remember that the Glauber-Sudarshan $P$-function
$P(\{\xi_{m}\},t)$ can be derived from the Wigner distribution by replacing
$\tilde{J}_{mn}$ by $J_{mn}$. Using such a $P$-function we obtain a compact
expression of the evolved density operator associated with the initial state
(\ref{42}), given by%
\begin{align}
\rho_{S}(t)  &  =\sum_{\jmath}%
{\displaystyle\sum\limits_{\left\{  x_{m}\right\}  }}
{\displaystyle\sum\limits_{\left\{  y_{m}\right\}  }}
p_{\jmath}\left(  C_{\left\{  y_{m}\right\}  }^{\left(  \jmath\right)
}\right)  ^{\ast}C_{\left\{  x_{m}\right\}  }^{\left(  \jmath\right)
}\nonumber\\
&  \times\left\{
{\displaystyle\prod\limits_{n}}
\sum_{q_{n}=0}^{x_{n}}\sum_{i_{n},j_{n},k_{n}}\frac{(-1)^{k_{n}}}{k_{n}%
!\sqrt{i_{n}!j_{n}!}}\frac{\sqrt{x_{n}!y_{n}!}}{q_{n}!\left(  x_{n}%
-q_{n}\right)  !\left(  y_{n}-x_{n}+q_{n}\right)  !}\right. \nonumber\\
&  \times\lim_{\varepsilon_{n}\rightarrow0}\frac{\partial^{i_{n}+j_{n}+2k_{n}%
}}{\partial\left(  \varepsilon_{n}\right)  ^{j_{n}+k_{n}}\partial\left(
\varepsilon_{n}^{\ast}\right)  ^{i_{n}+k_{n}}}\left[  \left(  \sum
_{m}\varepsilon_{m}\Theta_{mn}^{\ast}(t)\right)  ^{y_{n}-x_{n}+q_{n}}\right.
\nonumber\\
&  \times\left.  \left.  \left(  \sum_{m}\varepsilon_{m}^{\ast}\Theta
_{mn}(t)\right)  ^{q_{n}}\right]  \left\vert i_{n}\right\rangle \left\langle
j_{n}\right\vert \right\}  \exp\left(  \frac{1}{2}\sum_{m,n}\varepsilon
_{m}J_{mn}\left(  t\right)  \varepsilon_{n}^{\ast}\right)  \text{.} \label{48}%
\end{align}

By defining the parameters
\begin{subequations}
\label{50}%
\begin{align}
R_{\ell0}  &  =j_{\ell}+k_{\ell}\text{,}\label{50a}\\
S_{\ell0}  &  =i_{\ell}+k_{\ell}\text{,}\label{50b}\\
S_{\ell N}  &  =0\text{,}\label{50c}\\
\delta\left(  x\right)   &  =\left\{
\begin{array}
[c]{ccc}%
1 & \text{if} & x=0\\
0 & \text{if} & x\neq0
\end{array}
\right.  \text{,} \label{50d}%
\end{align}
we, alternatively, obtain the explicit form of the density operator
\end{subequations}
\begin{align}
\rho_{S}(t)  &  =\sum_{\jmath}%
{\displaystyle\sum\limits_{\left\{  x_{m}\right\}  }}
{\displaystyle\sum\limits_{\left\{  y_{m}\right\}  }}
p_{\jmath}\left(  C_{\left\{  y_{m}\right\}  }^{\left(  \jmath\right)
}\right)  ^{\ast}C_{\left\{  x_{m}\right\}  }^{\left(  \jmath\right)
}\left\{
{\displaystyle\prod\limits_{\ell}}
\sum_{q_{\ell}=0}^{x_{\ell}}\sum_{i_{\ell},j_{\ell},k_{\ell}=0}^{\infty}%
\frac{\sqrt{y_{\ell}!x_{\ell}!}}{\left(  x_{\ell}-q_{\ell}\right)  !}\right.
\nonumber\\
&  \times\frac{(-1)^{k_{\ell}}\left(  i_{\ell}+k_{\ell}\right)  !\left(
j_{\ell}+k_{\ell}\right)  !}{k_{\ell}!\sqrt{i_{\ell}!j_{\ell}!}}\left[
{\displaystyle\prod\limits_{\ell^{\prime}}}
\sum_{S_{\ell,\ell^{\prime}}=0}^{S_{\ell,\ell^{\prime}-1}}\sum_{K_{\ell
,\ell^{\prime}}=0}^{S_{\ell,\ell^{\prime}-1}-S_{\ell\ell^{\prime}}}%
\sum_{R_{\ell,\ell^{\prime}}=0}^{R_{\ell,\ell^{\prime}-1}}\left(  \frac{1}%
{2}J_{\ell^{\prime}\ell}\left(  t\right)  \right)  ^{S_{\ell,\ell^{\prime}%
-1}-S_{\ell,\ell^{\prime}}-K_{\ell,\ell^{\prime}}}\right. \nonumber\\
&  \times\left.  \frac{\left[  \Theta_{\ell\ell^{\prime}}(t)\right]
^{K_{\ell,\ell^{\prime}}}\left[  \Theta_{\ell\ell^{\prime}}^{\ast}(t)\right]
^{R_{\ell,\ell^{\prime}-1}-R_{\ell,\ell^{\prime}}}}{\left(  R_{\ell
,\ell^{\prime}-1}-R_{\ell,\ell^{\prime}}\right)  !\left(  S_{\ell,\ell
^{\prime}-1}-S_{\ell,\ell^{\prime}}-K_{\ell,\ell^{\prime}}\right)
!K_{\ell,\ell^{\prime}}!}\right]  \delta\left(  q_{\ell}-\sum_{\ell^{\prime}%
}K_{\ell^{\prime},\ell}\right) \nonumber\\
&  \times\left.  \delta\left[  R_{\ell,N}-\sum_{\ell^{\prime}}\left(
S_{\ell^{\prime},\ell-1}-S_{\ell^{\prime},\ell}-K_{\ell^{\prime},\ell}\right)
\right]  \delta\left[  y_{\ell}-x_{\ell}+q_{\ell}-\sum_{\ell^{\prime}}\left(
R_{\ell^{\prime},\ell-1}-R_{\ell^{\prime},\ell}\right)  \right]  \left\vert
i_{\ell}\right\rangle \left\langle j_{\ell}\right\vert \right\}  \text{.}
\label{49}%
\end{align}

For the case where all the reservoirs are at $0$ K, so that $J_{mn}=0$, only
the terms with $K_{\ell,\ell^{\prime}}=S_{\ell,\ell^{\prime}-1}-S_{\ell
,\ell^{\prime}}$ survive in the summation over $K_{\ell,\ell^{\prime}}$ of
expression (\ref{49}). Therefore, at $0$ K, the density operator (\ref{49})
reduces to the expression
\begin{align}
\rho_{S}(t) &  =\sum_{\jmath}%
{\displaystyle\sum\limits_{\left\{  x_{m}\right\}  }}
{\displaystyle\sum\limits_{\left\{  y_{m}\right\}  }}
p_{\jmath}\left(  C_{\left\{  y_{m}\right\}  }^{\left(  \jmath\right)
}\right)  ^{\ast}C_{\left\{  x_{m}\right\}  }^{\left(  \jmath\right)  }\left(
%
{\displaystyle\prod\limits_{\ell}}
\sum_{q_{\ell}=0}^{x_{\ell}}\frac{\sqrt{y_{\ell}!x_{\ell}!}}{\left(  x_{\ell
}-q_{\ell}\right)  !}\sum_{k_{\ell}=0}^{\infty}\frac{(-1)^{k_{\ell}}}{k_{\ell
}!}\right)  \nonumber\\
&  \times\left\vert \mathcal{F}(\left\{  q_{\ell}\right\}  ,\left\{  k_{\ell
}\right\}  ,t)\right\rangle \left\langle \mathcal{F}(\left\{  y_{\ell}%
-x_{\ell}+q_{\ell}\right\}  ,\left\{  k_{\ell}\right\}  ,t)\right\vert
\label{51}%
\end{align}
as already presented in Ref. \cite{MickelGeral}, where we have defined, with
$S_{\ell N}=0$, the superposition of product states
\begin{align}
\left\vert \mathcal{F}(\left\{  q_{\ell}\right\}  ,\left\{  k_{\ell}\right\}
,t)\right\rangle  &  =\bigotimes\limits_{\ell}\sum_{j_{\ell}=0}^{\infty}%
\frac{\left(  j_{\ell}+k_{\ell}\right)  !}{\sqrt{j_{\ell}!}}\left(
{\displaystyle\prod\limits_{\ell^{\prime}}}
\sum_{S_{\ell,\ell^{\prime}}=0}^{S_{\ell,\ell^{\prime}-1}}\frac{\left[
\Theta_{\ell\ell^{\prime}}(t)\right]  ^{S_{\ell,\ell^{\prime}-1}-S_{\ell
,\ell^{\prime}}}}{\left(  S_{\ell,\ell^{\prime}-1}-S_{\ell,\ell^{\prime}%
}\right)  !}\right)  \nonumber\\
&  \times\delta\left(  q_{\ell}-\sum_{\ell^{\prime}}\left(  S_{\ell^{\prime
},\ell-1}-S_{\ell^{\prime},\ell}\right)  \right)  \left\vert j_{\ell
}\right\rangle \text{.}\label{52}%
\end{align}

\textbf{Acknowledgments}

We wish to express thanks for the support from FAPESP and CNPq Brazilian
agencies .

\textbf{Figure captions}

Fig. 1 Sketch of a dissipative symmetric network of $N$ oscillators, where
each one interacts with each other, apart from its own reservoir.

Fig. 2 Sketch of a dissipative symmetric network of $N$ noninteracting
normal-mode oscillators, each one interacting with all the reservoirs.


\begin{thebibliography}{99}                                                                                               %


\bibitem {Feder}D. L. Feder, Phys. Rev. Lett. \textbf{97}, 180502 (2006).

\bibitem {Spin}M. Christandl, N. Datta, A. Ekert, and A. J. Landahl, Phys.
Rev. Lett. \textbf{92}, 187902 (2004); A. Kay, Phys. Rev. A \textbf{73},
032306 (2006); Phys. Rev. Lett. \textbf{98}, 010501 (2007).

\bibitem {Plenio}M. B. Plenio, J. Hartley, and J. Eisert, New J. Phys.
\textbf{6 }(2004).

\bibitem {Topology}V. Kostak, G. M. Nikolopoulos, I. Jex, Phys. Rev. A
\textbf{75}, 042319 (2007).

\bibitem {Bose}D. Burgarth and S, Bose, New Journal of Physics \textbf{7}, 135 (2005).

\bibitem {Brito}G. Burkard and F. Brito, Phys. Rev. B \textbf{72}, 054528 (2005).

\bibitem {Ficek}Z. Ficek and R. Tanas, Phys. Rep. \textbf{372}, 369 (2002).

\bibitem {Mickel1}M. A. de Ponte, M. C. de Oliveira, and M. H. Y. Moussa, Ann.
Phys (N.Y.). \textbf{317}, 72 (2005).

\bibitem {Mickel2}M. A. de Ponte, M. C. de Oliveira, and M. H. Y. Moussa,
Phys. Rev. A \textbf{70}, 022324 (2004); \textit{ibid.} \textbf{70}, 022325 (2004).

\bibitem {MickelDFS}M. A. de Ponte, S. S. Mizrahi, and M. H. Y. Moussa, Ann.
Phys. (N.Y.) \textbf{322}, 2077 (2007).

\bibitem {MickelGeral}M. A. de Ponte, S. S. Mizrahi, and M. H. Y. Moussa,
Phys. Rev. A \textbf{76}, 032101 (2007).

\bibitem {ZR}P. Zanardi and M. Rasetti, Phys. Rev. Lett. \textbf{79}, 3306
(1997); D. A. Lidar, I. L. Chuang, and K. B. Whaley, Phys. Rev. Lett.
\textbf{81}, 2594 (1998); E. Knill, R. Laflamme, and L. Viola, \textit{ibid}
\textbf{84}, 2525 (2000); A. R. Bosco de Magalh\~{a}es and M. C. Nemes, Phys.
Rev. A \textbf{70}, 053825 (2004); D. A. Lidar and K. B. Whaley, quant-ph/0301032.

\bibitem {Gardiner}C. W. Gardiner, \textit{Stochastic Methods}
(Springer-Verlag, Berlin,1983).

\bibitem {Mickel4}M. A. de Ponte, S. S. Mizrahi, and M. H. Y. Moussa, to be
published elsewhere.

\bibitem {livroSalomon}Roger A. Horn and Charles R. Johnson, \textit{Topics in
Matrix Analysis} (Cambridge University Press, New York, 1991).
\end{thebibliography}
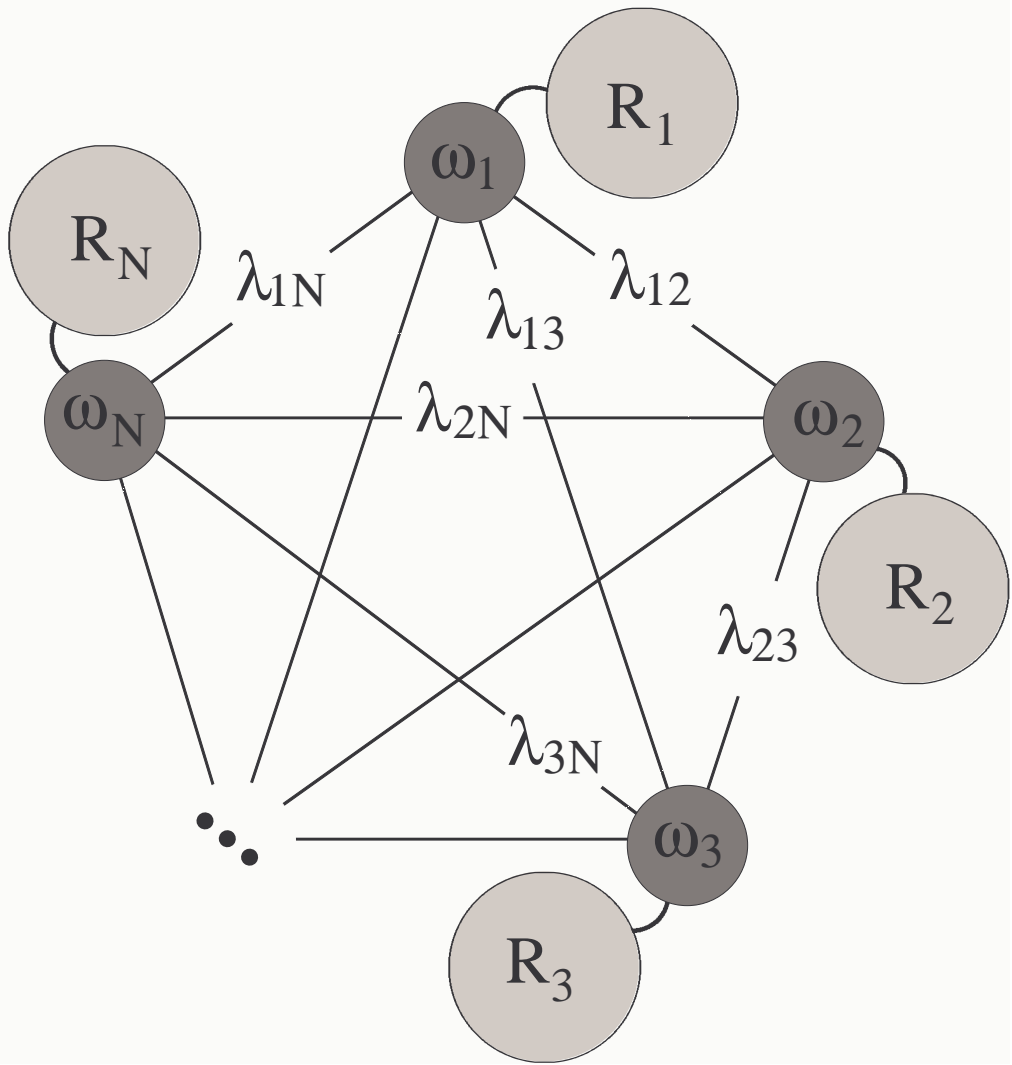
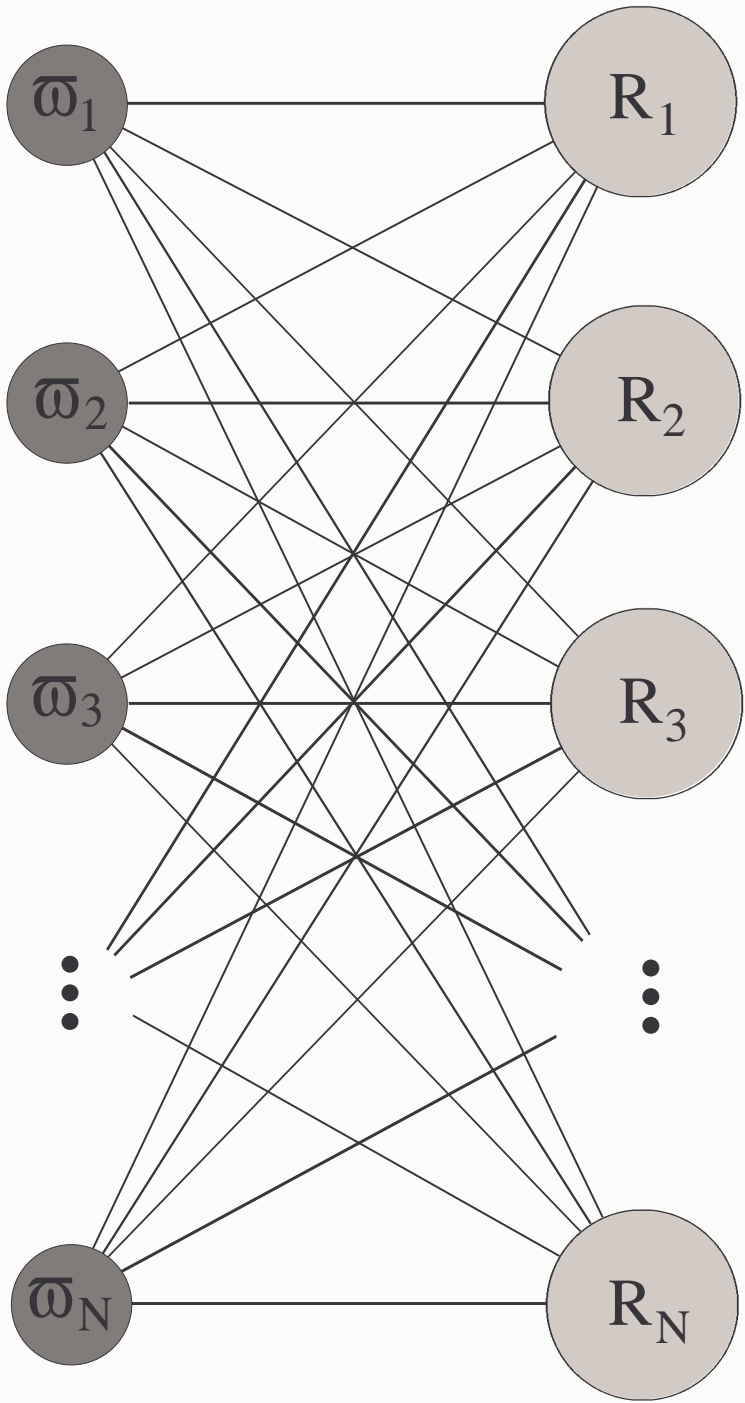
\end{document}